\documentclass[acmtocl,acmnow]{myacmtrans2m}

\newtheorem{theorem}{Theorem}[section]

\newtheorem{corollary}[theorem]{Corollary}
\newtheorem{proposition}[theorem]{Proposition}
\newtheorem{lemma}[theorem]{Lemma}
\newdef{definition}[theorem]{Definition}
\newdef{remark}[theorem]{Remark}
\newdef{example}[theorem]{Example}
\newdef{assumption}[theorem]{Assumption}
\newdef{algorithm}[theorem]{Algorithm}
\newtheorem{lemdefn}[theorem]{Lemma and Definition}

\usepackage{amsmath}
\usepackage{amssymb}
\usepackage{xspace}
\usepackage{bbold}
\usepackage{bbm}

\newcommand{\CoCASL}{\textsc{CoCasl}\xspace}

\newcommand{\Land}{\bigwedge}
\newcommand{\Lor}{\bigvee}
\newcommand{\cf}{\mathbbm{1}}

\newcommand{\Ext}{\mathcal{E}}
\newcommand{\Clause}{\mathsf{Cl}}

\newcommand{\Sum}{\mathcal{S}}

\newcommand{\mi}[1]{\mathit{#1}}

\newlength{\croutw}
\newlength{\crouth}
\newcommand{\crossout}[1]%
        {\settowidth{\croutw}{$#1$}\settoheight{\crouth}{$#1$}#1%
        \hspace{-1.0\croutw}\raisebox{0.3\crouth}{\rule{\croutw}{0.1ex}}}

\newcommand{\commentout}[1]{\ignorespaces}

\newcommand{\infrule}[2]{\frac{#1}{#2}}

\newcommand{\Pow}[1]{{\cal P}(#1)}

\newcommand{\bit}{\begin{itemize}}
\newcommand{\eit}{\end{itemize}}

        {\begin{enumerate}}%
        {\end{enumerate}}
        {\begin{enumerate}}%
        {\end{enumerate}}
        {\begin{enumerate}}%
        {\end{enumerate}}
        {\begin{enumerate}}%
        {\end{enumerate}}

\newcommand{\Cat}{\mathbf}

\newcommand{\Op}{{op}}

\newcommand{\lrule}[3]{(#1)\;\;\infrule{#2}{#3}}

\newcommand{\laxiom}[2]{(#1)\;\;#2}

\newcommand{\into}{\hookrightarrow}

\newcommand{\id}{{id}}

\newcommand{\powerset}{{\mathcal P}}
\newcommand{\finpowerset}{{\powerset_\omega}}

\newcommand{\argument}{\_\!\_}

\newcommand{\Set}{\Cat{Set}}

\newcommand{\Sem}[1]{{[\![#1]\!]}}

\newcommand{\conj}{\wedge}
\newcommand{\disj}{\vee}
\newcommand{\modimpl}{\to}
\newcommand{\modiff}{\leftrightarrow}

\newcommand{\entails}{\vdash}

        {\smallskip\par\noindent\qquad$\begin{array}{@{}l@{{}={}}l}}%
        {\end{array}$\par\noindent}

\newenvironment{axarray}%
	       {\begin{array}{@{\hspace{5em}}p{5em}p{50em}}
	       }{\end{array}}

\newlength{\myboxwidth}
\newcommand{\Lang}{\mathcal{L}}	
\newcommand{\FLang}{\mathcal{F}}

\newcommand{\plbox}[1]{[#1]}
\newcommand{\pldiamond}[1]{\langle#1\rangle}
\newcommand{\gldiamond}[1]{\Diamond_{#1}}
\newcommand{\glbang}[1]{\Diamond!_{#1}}
\newcommand{\glbox}[1]{\square_{#1}}
\newcommand{\coalbox}[1]{[#1]}

\newcommand{\HMLBox}[1]{\square\hspace{-5.5pt}\raisebox{2pt}{$\scriptscriptstyle{#1}$}\;}

\newcommand{\Nat}{{\mathbb{N}}}
\newcommand{\Int}{{\mathbb{Z}}}
\newcommand{\Rat}{{\mathbb{Q}}}
\newcommand{\Real}{{\mathbb{R}}}

\newcommand{\PDist}{D_\omega}

\newcommand{\Bag}{\mathcal{B}}
\newcommand{\Prop}{\mathsf{Prop}}
\newcommand{\Up}{\mathsf{Up}}

\newcommand{\satisfies}{\vDash}

\newcommand{\Rules}{\mathcal{R}}
\newcommand{\RulesC}{\Rules_C}

\newcommand{\UpP}{\mathsf{Up}\mathcal{P}}

\newcommand{\sgn}{\mathit{sgn}}
\newcommand{\size}{\mathit{size}}

\newcommand{\ModAts}{\mathit{MA}}

\newcommand{\PLentails}{\entails_{\mi{PL}}}

\newcommand{\CB}{\mathcal{B}}
\newcommand{\CA}{\mathcal{A}}

\newcommand{\Kripke}{\mathcal{K}}
\newcommand{\ModSig}{\Lambda}
\newcommand{\contrapow}{\mathcal{Q}}
\newcommand{\GML}{\mi{GML}}
\newcommand{\PML}{\mi{PML}}
\newcommand{\Coal}{\mi{Coal}}

\newcommand{\NP}{\mi{NP}}
\newcommand{\PSPACE}{\mi{PSPACE}}
\newcommand{\APTIME}{\mi{APTIME}}

\newcommand{\eat}[1]{}

\title{PSPACE Bounds for Rank-1 Modal Logics}

\author{LUTZ SCHR{\"O}DER\\
DFKI-Lab Bremen and Dept.\ of Comput.\ Sci., Universit\"at Bremen
 \and 
DIRK PATTINSON\\
Department of Computing, 
Imperial College London}

\begin{abstract}
For lack of general algorithmic methods that apply to wide classes of
logics, establishing a complexity bound for a given modal logic is often
a laborious task. The present work is a step towards a general theory of
the complexity of modal logics.  Our main result is that all
\mbox{rank-1} logics enjoy a shallow model property and thus are, under
mild assumptions on the format of their axiomatisation, in $\PSPACE$.
This leads to a unified derivation of tight $\PSPACE$-bounds for a
number of logics including $K$, $KD$, coalition logic, graded modal logic,
majority logic, and probabilistic modal logic. Our generic algorithm
moreover finds tableau proofs that witness pleasant proof-theoretic
properties including a weak subformula property. This generality is made
possible by a coalgebraic semantics, which conveniently abstracts from
the details of a given model class and thus allows covering a broad
range of logics in a uniform way.
\end{abstract}

\category{F.2.2.}{Analysis of Algorithms and Problem
  Complexity}{Nonnumerical Algorithms and Problems}[Complexity of Proof
  Procedures]



\begin{document}

            
\maketitle

\runningfoot{}
\firstfoot{}

\section{Introduction}

Modal logics are attractive from a computational point of view, as they
often combine expressiveness with decidability.  For many modal logics
not involving dynamic features, satisfiability is known to be in
$\PSPACE$. This is typically proved for one logic at a time, e.g.\
by modifications of the witness algorithm for the modal
logic~$K$~\cite{Ladner77,BlackburnEA01}, but also using markedly different
methods such as the constraint-based $\PSPACE$-algorithm for
graded modal logic~\cite{Tobies01}. \citeN{Vardi89} gives a first
glimpse of a generalisable method, equipping various epistemic logics
with a neighbourhood frame semantics and showing them to be in $\NP$
and $\PSPACE$, respectively (with the $K$ axiom being responsible
for $\PSPACE$-hardness; recent work by~\citeN{HalpernRego06} shows that
negative introspection brings the complexity back down to
$\NP$). Nevertheless, there is to date no generally applicable
theorem that allows establishing $\PSPACE$-bounds for large classes
of modal logics in a uniform way.

Here, we generalise the methods of~\cite{Vardi89} to obtain $\PSPACE$
bounds for rank-1 modal logics, i.e.\  logics axiomatisable by formulas whose
modal depth uniformly equals one, in a systematic
way. Although presently limited to rank~1, our approach covers numerous
relevant and non-trivial examples. We recover known
$\PSPACE$ bounds not only for normal modal logics such as $K$ and
$KD$, but most notably also for a range of non-normal modal logics such
as graded modal logic~\cite{Fine72}, coalition logic~\cite{Pauly02},
and probabilistic modal
logic~\cite{LarsenSkou91,HeifetzMongin01}. Moreover, our methods 
lead
to a previously unknown $\mi{PSPACE}$ upper bound for majority
logic~\cite{PacuitSalame04} 
that was independently discovered by~\citeN{DemriLugiez06} at the
same time.
These logics are far from exotic: graded
modal logic plays a role e.g.\ in decision support and knowledge
representation~\cite{vdHoekMeyer92,OhlbachKoehler99}, and probabilistic
modal logic has appeared in connection with model
checking~\cite{LarsenSkou91} and in modelling economic
behaviour~\cite{HeifetzMongin01}.

The key to such a degree of generality is to parametrise the theory over
the type of systems defining the semantics, using coalgebraic methods.
Coalgebra conveniently abstracts from the details of a concrete class of
models as it encapsulates
the precise nature of models 
in an endofunctor on the category of sets. As specific instances, one
obtains e.g.\ (serial) Kripke frames, (monotone) neighbourhood
frames~\cite{HansenKupke04}, game frames~\cite{Pauly02}, probabilistic
transition systems and automata~\cite{Rabin63,BartelsEA04}, weighted
automata, linear automata~\cite{CarlylePaz71}, and
multigraphs~\cite{DAgostinoVisser02}. Despite the broad range of systems
covered by the coalgebraic approach, a substantial body of concepts and
non-trivial results has emerged, encompassing e.g.\ generic notions of
bisimilarity and coinduction~\cite{Bartels03},
corecursion~\cite{TuriPlotkin97}, duality, and ultrafilter
extensions~\cite{KupkeEA05}. On the applications side, coalgebraic modal
logic  features in actual specification languages such as
the object oriented specification language CCSL~\cite{RotheTewsJacobs01}
and \CoCASL~\cite{MossakowskiEA04}.

The coalgebraic study of computational aspects of modal logic was
initiated in~\cite{Schroder06}, where the finite model property and
associated $\mi{NEXPTIME}$-bounds were proved.  Here, we push these
results further and present a shallow model property based on
coalgebraic semantics.  This leads to a generic $\PSPACE$-algorithm for
deciding satisfiability that traverses a shallow model and strips off
one layer of modalities in every step. Alternatively, our algorithm may
be seen as computing a shallow proof that enjoys a number of pleasant
proof-theoretic properties, including a weak subformula property (i.e.\
it mentions only propositional combinations of subformulas of the goal).

The model construction relies on extending the axiomatisation of a given
logic to a set of rules which is closed under \emph{rule resolution},
i.e.\ every resolvent of two substituted rule conclusions can also be
derived directly using a third rule.  This process typically results in
an infinite but recursive set of rules. Resolution closedness then
enables us to build the shallow model using induction on the modal depth
of formulas. Since we are working with an infinite set of rules, we have
to impose a second condition
to ensure that we can decide satisfiability: a rule set is closed under
\emph{contraction} if every substituted rule conclusion with duplicate
literals can be derived using a substitution instance of a second rule
in whose conclusion all literals remain distinct. The decision procedure
will run in $\PSPACE$ if both closure under resolution and closure under
contraction can be controlled, i.e.\ there is a polynomial bound on the
size of rules that are applicable at every step of the
deductive process. This turns out to be the case for all examples
mentioned above.

The material is organised as follows. In Section~\ref{sec:prelim}, we
give a brief introduction to the generic coalgebraic semantics of modal
logic. In Section~\ref{sec:deduction}, we discuss deduction systems for
coalgebraic modal logics and their properties, notably the (equivalent)
central notions of strict one-step completeness and reduction closedness
of rule sets. Sections~\ref{sec:shallowmodels}
and~\ref{sec:shallowproofs} are devoted to the tableau-based shallow
model construction and the proof-theoretic view thereof. The ensuing
$\mi{PSPACE}$-algorithm and its example applications are presented in
Section~\ref{sec:pspace}.

\section{Coalgebraic Modal Logic}\label{sec:prelim}

We briefly recapitulate the basics of the coalgebraic interpretation of
modal logic. To begin, we fix the syntactic framework. A \emph{modal
signature} is just a set~$\ModSig$ of unary modal operators (all our
results generalise straightforwardly to a polyadic setting as in
\cite{Schroder05}).  The signature $\ModSig$ induces a modal
language~$\FLang(\ModSig)$, with formulas $\phi,\psi\in\FLang(\ModSig)$
defined by the grammar
\begin{equation*}
\phi :: =  \bot\ \mid\ \phi\conj\psi\ \mid\ \neg\phi\ \mid\ 
L \phi,
\end{equation*}
where~$L$ ranges over~$\ModSig$.  Disjunctions $\phi\disj\psi$, truth
$\top$, and other boolean operations are defined as usual. The
\emph{depth} of a formula is its maximal nesting depth of modal
operators. 


We work in the framework of \emph{coalgebraic modal logic},
introduced by
\citeN{Pattinson04}, generalising previous
results~\cite{Jacobs00,Roessiger00,Kurz01,Pattinson01},
where modal languages are interpreted over coalgebras for a
$\Set$-functor:
\begin{definition}\label{def:coalg}
\cite{Rutten00} Let $T: \Set \to \Set$ be a functor, referred to as the
\emph{signature functor}, where $\Set$ is the category of sets. A
\emph{$T$-coalgebra}~$A=(X,\xi)$ is a pair ($X, \xi)$ where $X$ is a set
(of \emph{states}) and $\xi: X \to TX$ is a function called the
\emph{transition} function. A \emph{morphism} $f:A\to B$ between
$T$-coalgebras $A=(X,\xi)$ and $B=(Y,\zeta)$ is a map $f:X\to Y$ such
that $Tf\xi=\zeta f$. 
\end{definition}

We view coalgebras as generalised transition systems: the transition
function delivers a structured set of successors and observations for a
state.  Mutatis mutandis, we can in fact allow $T$ to take proper
classes as values, as we never iterate $T$ or otherwise assume that
$TX$
is a set; details are left implicit. This allows us to treat more
examples, in particular Pauly's coalition logic (Example
\ref{expl:coalgml}.\ref{expl:coalitionlogic} below).
\begin{assumption}\label{ass:injective}
We can assume w.l.o.g.\ that $T$ preserves injective maps~\cite{Barr93}.
For convenience of notation, we will in fact sometimes assume that $TX
\subseteq TY$ in case $X \subseteq Y$. Moreover, we assume w.l.o.g.\
that $T$ is non-trivial, i.e.\ $TX=\emptyset\implies X=\emptyset$
(otherwise, $TX=\emptyset$ for all $X$).
\end{assumption}
\begin{definition}\label{def:subcoalg}
If for a subset $Z\subseteq X$ of a coalgebra
$A=(X,\xi)$, $\xi$ restricts to a map $\xi_Z:Z\to TZ$, then $C=(Z,\xi_Z)$
is a \emph{subcoalgbra} of $A$; in this case, the
inclusion $Z\into X$ is a morphism $C\to A$.
\end{definition}

In the same way that the signature functor abstracts from a concrete
class of models, the interpretation of modal operators is encapsulated
in terms of predicate liftings:
\begin{definition}
A \emph{predicate lifting} for a functor~$T$ is a natural
transformation
\begin{equation*}
\contrapow \to \contrapow\circ T^\Op,
\end{equation*} 
where~$\contrapow$ denotes the contravariant powerset functor
$\Set^\Op\to\Set$ (i.e.\ $\contrapow(X)=\powerset(X)$ is the powerset,
and $\contrapow f(B)=f^{-1}[B]$ for $f:X\to Y$ and $B\in\contrapow(X)$).
\end{definition}
A coalgebraic semantics for a modal signature $\ModSig$ is given by a
\emph{$\ModSig$-structure}, consisting of a signature functor~$T$ and an
assignment of a predicate lifting~$\Sem{L}$ for~$T$ to every modal
operator~$L\in\ModSig$; by abuse of notation, we refer to the entire
$\ModSig$-structure just as~$T$. Given a $\ModSig$-structure~$T$, the
satisfaction relation~$\models_C$ between states~$x$ of a $T$-coalgebra
$C = (X, \xi)$ and $\FLang(\ModSig)$-formulas is defined inductively,
with the usual clauses for the boolean operations.  The clause for the
modal operator~$L$ is
\begin{equation*}
  x \models_C L \phi \iff
  \xi(x) \in \Sem{L}_C(\Sem{\phi}_C),
\end{equation*}
where $\Sem{\phi}_C=\{ x \in X \mid x \models_C \phi\}$.  We drop the
subscripts~$C$ when these are clear from the context.

We occasionally make use of the fact that the logic $\FLang(\ModSig)$ is
\emph{adequate} for $T$-coalgebras~\cite{Pattinson04}:
\begin{proposition}\label{prop:adequacy}
If $f:A\to B$ is a morphism of $T$-coalgebras, then
\begin{equation*}
x\models_A\phi\quad\textrm{iff}\quad f(x)\models_B\phi
\end{equation*}
for all states $x$ in $A$ and all $\FLang(\Lambda)$-formulas $\phi$.
\end{proposition}


%
Our main interest here is in the local \emph{satisfiability problem}:
\begin{definition}\label{def:lang}
An $\FLang(\ModSig)$-formula~$\phi$ is \emph{satisfiable} (over~$T$) if
there exist a $T$-coalgebra $A = (X, \xi)$ and a state~$x$ in~$X$ such that
$x\models_A\phi$. Dually,~$\phi$ is \emph{valid} if~$x\models_A\phi$ for
all $T$-coalgebras $A = (X, \xi)$ and all $x \in X$.
\end{definition}
\begin{example}\label{expl:coalgml}
\cite{Pattinson04,CirsteaPattinson07,Schroder06} We illustrate how the
coalgebraic approach subsumes a large class of modal logics. This
includes not only logics with a standard Kripke semantics, but in
particular also non-normal modal logics whose semantics is defined over
structures that differ substantially from classical Kripke frames.
\begin{longenum}
\item\label{expl:power-functor} \emph{Modal logic $K$:} The signature
  $\ModSig_K$ of the modal logic~$K$ consists of a single modal
  operator~$\Box$.  Let~$\powerset$ be the covariant powerset
  functor. Then $\powerset$-coalgebras are graphs, thought of as
  transition systems or indeed Kripke frames. A $\ModSig_K$-structure
  over~$\powerset$ is defined by
  \begin{equation*}
    \Sem{\Box}_X(A)=\{B\in\powerset(X)\mid B\subseteq A\};
  \end{equation*}
  this induces precisely the standard Kripke semantics of modal
  logic (note that no restrictions are imposed on frames). 
\item\label{expl:kd} \emph{Modal logic $KD$:} $KD$ is obtained from~$K$
  by adding the axiom~$\neg\Box\bot$, i.e.\ by restricting the semantics
  to \emph{serial} Kripke frames~$(X,R)$, characterized by the condition
  that for every state~$x$, there exists a state~$y$ such that
  $xRy$. Thus, the signature~$\ModSig_{KD}$ of the normal modal logic
  $KD$ is the same as that of~$K$, and a $\ModSig_{KD}$-structure is
  defined in the same way as for~$K$, but over the non-empty powerset
  functor~$\powerset^*$ defined by
  $\powerset^*(X)=\{A\in\powerset(X)\mid A\neq\emptyset\}$.
\item \label{expl:frame-functor}\emph{Modal logic $E$:} The signature
  $\ModSig_E$ of the modal logic~$E$, the smallest classical modal
  logic~\cite{Chellas80}, has a single modal operator~$\Box$; the proof
  system of $E$ comprises, besides propositional reasoning, only
  replacement of equivalents (i.e.\ the rule $a\modiff b/\Box a\modimpl
  \Box b$). The standard neighbourhood semantics of~$E$
  is coalgebraically captured by a $\ModSig_E$-structure over the
  \emph{neighbourhood functor} $N = \contrapow \circ \contrapow^\Op$
  (composition of the contravariant powerset functor with itself);
  coalgebras for this functor are neighbourhood frames.  The modal
  operator~$\Box$ is interpreted over~$N$ by
  \begin{equation*}
    \Sem{\Box}_X(A)=\{\alpha\in N(X)\mid A\in\alpha\}.
  \end{equation*}
\item \label{expl:monotonicity} \emph{Modal logic $M$:} The modal
  logic~$M$, the smallest monotonic modal logic~\cite{Chellas80}, is
  obtained from the modal logic~$E$ by adding the monotonicity rule
  $a\modimpl b/\Box a\modimpl \Box b$. The neighbourhood semantics
  of~$M$ is captured coalgebraically analogously to the previous example
  as a structure over the subfunctor~$\UpP$ of $N$ assigning to a
  set~$X$ the set of upwards closed subsets of $\contrapow
  X$. Coalgebras for~$\UpP$ are monotone neighbourhood
  frames~\cite{HansenKupke04}.
\item \label{expl:bag} \emph{Graded modal logic}~\cite{Fine72}: The
  modal signature of \emph{graded modal logic} (GML) is
  $\ModSig_\GML=\{\gldiamond{k}\mid k\in\Nat\}$; the intended reading of
  $\gldiamond{k}\phi$ is `$\phi$ holds in more than~$k$ successor
  states'. The semantics of GML is originally defined by counting
  successor states in Kripke frames. This semantics fails to be
  coalgebraic, as the naturality condition for the associated predicate
  liftings fails. However, one may define a coalgebraic semantics which
  is equivalent for purposes of satisfiability~\cite{Schroder06}, as
  follows. The \emph{finite multiset} (or \emph{bag}) functor~$\Bag$
  maps a set~$X$ to the set of maps $B:X\to\Nat$ with finite support,
  the intuition being that~$B$ is a multiset containing $x\in X$ with
  multiplicity~$B(x)$.  We extend~$B$ to~$\powerset(X)$ by putting
  $B(A)=\sum_{x\in A}B(x)$.  The action on morphisms $f: X \to Y$ is
  then given by $\Bag f: \Bag X \to \Bag Y, B \mapsto\lambda y.\,
  B(f^{-1}[\{y\}])$.  Coalgebras for~$\Bag$ are directed graphs with
  $\Nat$-weighted edges, often referred to as
  \emph{multigraphs}~\cite{DAgostinoVisser02}. The graded modal operator
  $\gldiamond{k}$ is intepreted over~$\Bag$ by
  \begin{equation*}
    \textstyle
    \Sem{\gldiamond{k}}_X(A)=\{B: X \to \Nat \in \Bag(X) \mid  B(A) > k\}.
  \end{equation*}
  Thus, $x\satisfies \gldiamond{k}\phi$ for a state~$x$ in a
  $\Bag$-coalgebra iff~$\phi$ holds for more than~$k$ successor states
  of~$x$, taking into account multiplicities.  

  The dual operators~$\neg\gldiamond{k}\neg$ are denoted~$\glbox{k}$,
  i.e.~$\glbox{k}\phi$ reads `$\phi$ fails in at most~$k$ successor
  states'. Note that~$\glbox{k}$ is monotone, but fails to be normal
  unless $k=0$. A non-monotone variation of GML arises when negative
  multiplicities are admitted.
\item\label{expl:majority} \emph{Majority logic}~\cite{PacuitSalame04}:
   Graded modal logic is extended to \emph{majority logic} by adding a
   \emph{weak majority operator}~$W$, read `in at least half of the
   successor states, it is the case that \dots'. The structure for GML
   over the multiset functor~$\Bag$ described in the previous example is
   extended to~$W$ by putting
  \begin{equation*}
    \Sem{W}_X(A)=\{B:X\to\Nat\in\Bag(X)\mid\textstyle B(A)\ge
    B(X-A)\}.
  \end{equation*}
  The dual operator $M=\neg W\neg$ captures strict majority `in more
  than half of the successor states, it is the case that'.
\item\label{expl:distr} \emph{Probabilistic modal
  logic}~\cite{LarsenSkou91,HeifetzMongin01}: The modal
  signature~$\ModSig_\PML$ of \emph{probabilistic modal logic} (PML)
  comprises operators~$L_p$, $p\in[0,1]\cap\Rat$, to be read `in the
  next step, it is with probability at least~$p$ the case that\dots'. We
  define a $\ModSig_\PML$-structure over the \emph{finite distribution
  functor}~$\PDist$ which maps a set~$X$ to the set of probability
  distributions on~$X$ with finite support. Coalgebras for $\PDist$ are
  probabilistic transition systems (also called \emph{probabilistic type
  spaces}~\cite{HeifetzMongin01}) with finite branching degree. Our
  definition contrasts with that of~\cite{HeifetzMongin01}, where there
  is no restriction on the branching degree, but since PML has the
  finite model property (cf.\ loc.\ cit.), this has no bearing on
  satisfiability. The interpretation of $L_p$ over~$\PDist$ is defined
  by
  \begin{equation*}
  \Sem{L_p}(A)=\{P\in\PDist X\mid PA\ge p\}.
  \end{equation*}
  PML is non-normal ($L_p (a\disj b) \modimpl L_p a\disj L_p b$ is not
  valid for $p>0$).
\item\label{expl:coalitionlogic} \emph{Coalition logic}~\cite{Pauly02}:
  Let $N=\{1,\dots,n\}$ be a fixed set of \emph{agents}. Subsets of~$N$
  are called \emph{coalitions}. The signature~$\ModSig_{\Coal}$ of
  coalition logic consists of modal operators~$\coalbox{C}$, where~$C$
  ranges over coalitions, read `coalition~$C$ has a collaborative
  strategy to ensure that \dots'. A coalgebraic semantics for coalition logic
  is based on the class-valued signature functor~$T$ defined by
  \begin{equation*}
    TX = \lbrace (S_1, \dots, S_n, f) 
    \mid\textstyle \emptyset\neq S_i \in \Set,
    f: \prod_{i\in N} S_i \to X
    \rbrace.
  \end{equation*}
  The elements of~$TX$ are understood as \emph{strategic games} with set
  $X$ of states, i.e.\ tuples consisting of nonempty sets~$S_i$ of
  \emph{strategies} for all agents~$i$, and an \emph{outcome function}
  $(\prod S_i)\to X$. A~$T$-coalgebra is a \emph{game
  frame}~\cite{Pauly02}.
  We denote the set $\prod_{i\in C}S_i$ by~$S_C$, and for $\sigma_C\in
  S_C, \sigma_{\bar C}\in S_{\bar C}$, where $\bar C=N-C$,
  $(\sigma_C,\sigma_{\bar C})$ denotes the obvious element of $\prod_{i\in
  N} S_i$. A $\ModSig_\Coal$-structure over~$T$ is then defined by
  \begin{equation*}
    \Sem{\coalbox{C}}_X(A)=\{(S_1,\dots, S_n,f)\in TX\mid \exists \sigma_C\in
    S_C.\,\forall \sigma_{\bar C}\in S_{\bar C}.\,
    f(\sigma_C,\sigma_{\bar C})\in A\}.
  \end{equation*}
\end{longenum}
\end{example}

All the above examples can be canonically extended to systems that
process inputs from a set~$I$ by passing from the signature functor~$T$
to one of the functors~$T^I$ or $T(I\times \argument)$ and suitably
indexing the modal operators. We refer to \cite{CirsteaPattinson07} for
a detailed account of the induced logics.

\begin{remark}
  \label{rem:propsymb} In the modal grammar given above, atomic
  propositional symbols are deliberately not included. This is for the
  sake of both generality, as some modal logics such as Hennessy-Milner
  logic do not include such atomic propositions, and economy of
  presentation, as a set~$U$ of atomic propositional symbols may be
  integrated in the basic framework as follows. Given a modal signature
  $\ModSig$ and a $\ModSig$-structure~$T$, we define a structure for the
  modal signature $\ModSig_U=\ModSig\cup U$ over the functor~$T_U$
  defined by $T_U X = TX \times \powerset(U)$: modal
  operators from~$\ModSig$ are interpreted by taking the preimage of
  their interpretation over~$T$ under the projection $T_U\to T$, and a
  propositional symbol $a\in U$ is interpreted by putting
  \begin{equation*}
    \Sem{a}_X(A)=\{(t,B)\in TX\times\powerset(U)\mid a\in B\}.
  \end{equation*}
  Since~$\Sem{a}$ is independent of its argument, the modal operator~$a$
  can be written as just the propositional symbol~$a$ (without an
  argument formula). In a framework with polyadic modal
  operators~\cite{Schroder05}, propositional constants correspond to
  nullary modalities.  Some of the logics above indeed \emph{require}
  propositional symbols lest they collapse into triviality. This holds
  in those cases where $T1$ (for~$1$ a singleton set) is a singleton,
  e.g.\ probabilistic modal logic, coalition logic, and the modal
  logic~$\mi{KD}$. We nevertheless generally continue to omit the
  treatment of propositional symbols in the sequel, since the addition
  of propositional symbols as indicated above has no bearing on the rule
  sets forming the core of our method, and the model construction
  is entirely analogous.
\end{remark}

\section{Proof Systems For Coalgebraic Modal Logic}\label{sec:deduction}

\noindent
Our decision procedure for rank-1 logics relies on a complete
axiomatisation in a certain format. Deduction for modal logics with
coalgebraic semantics has been considered
in~\cite{Pattinson03,CirsteaPattinson07,KupkeEA05,Schroder06}. It has
been shown that every modal logic over coalgebras can be axiomatised in
rank~$1$ using either rank-$1$ axioms or rules leading from rank~$0$ to
rank~$1$~\cite{Schroder06}, essentially because functors, as opposed to
comonads, only encode the one-step behaviour of systems. Here, we focus
on rules. The crucial ingredients for the shallow model construction and
the ensuing $\PSPACE$ algorithm are novel notions of
\emph{resolution closure} and \emph{strict one-step completeness} of
rule sets.

For the remainder of the paper, we fix a modal signature
$\ModSig$ and a $\ModSig$-structure~$T$. We recall a few basic notions from
propositional logic, as well as notation for coalgebraic modal logic
introduced in~\cite{Pattinson03,CirsteaPattinson07}:
\begin{definition}\label{def:propstuff}
We denote the set of propositional formulas over a set~$V$ (consisting
e.g.\ of propositional variables or modal formulas) by $\Prop(V)$. Here,
we regard~$\neg$ and~$\conj$ as the basic connectives, with all other
connectives defined in the standard way. For $\phi,\psi\in\Prop(V)$, we
say that~$\phi$ \emph{propositionally entails}~$\psi$ and write
$\phi\PLentails\psi$ if $\phi\modimpl\psi$ is a propositional
tautology. Similarly, $\Phi\subseteq\Prop(V)$ propositionally entails
$\psi$ ($\Phi\PLentails\psi$) if there exist
$\phi_1,\dots,\phi_n\in\Phi$ such that
$\phi_1\land\dots\land\phi_n\PLentails\psi$.

A \emph{literal} over~$V$ is either an element of~$V$ or the negation of
such an element. We use the meta-variable~$\epsilon$ (possibly indexed)
to denote either nothing or~$\neg$, so that a literal over~$V$ has the
general form $\epsilon a$, $a\in V$. A \emph{clause} is a finite
(possibly empty) disjunction of literals, which then takes the form
$\Lor_{i=1}^n\epsilon_i a_i$ with $a_1, \dots, a_n \in V$.  Similarly, a
\emph{conjunctive clause} is a finite conjunction of literals. A
(conjunctive) clause is \emph{contracted} if all its literals are
distinct. The set of all clauses over~$V$ is denoted by
$\Clause(V)$. Although we regard clauses as formulas rather than sets of
literals, we shall sometimes use terminology such as `a literal is
contained in a clause' or `a clause contains another', with the obvious
meaning.  We denote by $\Up(V)$ the set $\{L a\mid L\in\ModSig,a\in
V\}$.

If~$V$ consists of propositional variables, then we have the usual
notions of valuation and substitution: A \emph{valuation} is just a map
$\kappa:V\to\{\top,\bot\}$ assigning boolean truth values to variables;
for $\phi\in\Prop(V)$, we write $\kappa\models\phi$ if~$\kappa$ is a
satisfying valuation for~$\phi$. More generally, given a set~$X$, a
\emph{$\powerset(X)$-valuation} for~$V$ is a map $V\to\powerset(X)$. For
$\phi\in\Prop(V)$, a $\powerset(X)$-valuation~$\tau$ induces in the
obvious way a subset $\Sem{\phi}\tau$ of~$X$; we write
$X,\tau\models\phi$ if $\Sem{\phi}\tau=X$.
Using the structure~$T$ for~$\Lambda$, we interpret
$\psi\in\Prop(\Up(V))$ as a subset $\Sem{\psi}\tau$ of~$TX$ by putting
$\Sem{L\phi}\tau=\Sem{L}\Sem{\phi}\tau$, and we write
$TX,\tau\models\psi$ if $\Sem{\psi}\tau=TX$.  Moreover, given a set~$Z$,
a \emph{$Z$-substitution} for~$V$ is a map $\sigma:V\to Z$; for a
formula~$\phi$ over~$V$ (e.g.\ $\phi\in\Prop(\Up(\Prop(V)))$), we denote
the result of performing the substitution~$\sigma$ on~$\phi$
by~$\phi\sigma$ and refer to $\phi\sigma$ as a \emph{$Z$-instance}
of~$\phi$.
\end{definition}

\begin{lemma}\label{lem:clause-entail}
For $\phi,\psi\in\Clause(V)$, $\phi\PLentails\psi$ iff either~$\phi$ is
contained in~$\psi$ or $\psi$ is a tautology (i.e.\ contains both $a$
and $\neg a$ for some $a\in V$).\qed
\end{lemma}
\begin{definition}
A \emph{(one-step) rule}~$R$ over a set~$V$ of propositional variables
is a rule~$\phi/\psi$, where $\phi\in\Prop(V)$
and~$\psi\in\Clause(\Up(V))$. We silently identify rules modulo
$\alpha$-equivalence. The rule~$R$ is \emph{sound} if, whenever
$\phi\sigma$ is valid for an $\FLang(\ModSig)$-substitution~$\sigma$,
then~$\psi\sigma$ is valid. Moreover,~$R$ is \emph{one-step sound} if
$TX,\tau\models\psi$ for each set~$X$ and each $\powerset(X)$-valuation
$\tau$ such that $X,\tau\models\phi$.
\end{definition}

Our hitherto informal use of the term \emph{rank-$1$ logic} formally
means \emph{axiomatisable by one-step rules}.  The term rank-1 logic has
been used in the
literature~\cite{Pattinson03,CirsteaPattinson07,KupkeEA05,Schroder06} to
describe logics axiomatisable by \emph{rank-1 axioms}, i.e.\ 
propositional combinations of formulas~$L \phi$ where~$L$ is a modal
operator and~$\phi$ is purely propositional (in the notation introduced
above, formulas from $\Prop(\Up(\Prop(V)))$). This class of axioms
includes e.g.\ the~$K$ axiom $\Box (a \to b) \to (\Box a \to \Box b)$,
but excludes axioms containing nested modalities or top-level
propositional variables such as the axioms~$4$ and~$T$, respectively. It
has been shown in~\cite{Schroder06} that one-step rules and rank-1
axioms determine the same class of logics.

\begin{remark}\label{rem:elimination}
We can always assume that every propositional variable~$a$ appearing in
the premise~$\phi$ of a one-step rule appears also in the conclusion:
otherwise, we can eliminate~$a$ by passing from~$\phi$ to
$\phi[\top/a]\disj\phi[\bot/a]$.
\end{remark}
\begin{proposition}\cite{Schroder06}
Every one-step sound rule is sound.\qed
\end{proposition}

The converse holds under additional assumptions~\cite{Schroder05}; note
however that the obviously sound rule $\bot/\bot$ is one-step sound iff
$T\emptyset=\emptyset$ (as is the case e.g.\ for PML).

A given set~$\Rules$ of one-step sound rules induces a proof system
for $\FLang(\ModSig)$ as follows.
\begin{definition}\label{def:provable}
Let~$\RulesC$ denote the set of rules obtained by extending~$\Rules$
with the \emph{congruence rule} 
\begin{equation*}
\lrule{C}{a\modiff b}{La\modimpl Lb}
\end{equation*}
for every $L\in\ModSig$.
(This rule of course implies a rule where~$\modimpl$ is replaced by
$\modiff$, which however does not fit the format for one-step rules.)
The set of \emph{provable} formulas is the smallest set closed under
propositional entailment and the rules in~$\RulesC$, with propositional
variables instantiated to formulas in $\FLang(\ModSig)$. We say that a
formula~$\phi$ is \emph{consistent} if~$\neg\phi$ is not provable.
\end{definition}
It is easy to see that this proof system is sound. Completeness requires
`enough' rules in the following sense.

\begin{definition}\label{def:reflexive}
The set~$\Rules$ is \emph{(strictly) one-step complete} if, whenever
$TX,\tau\models\chi$ for a set~$X$, $\chi\in\Clause(\Up(V))$, and a
$\powerset(X)$-valuation~$\tau$, then~$\chi$ is \emph{(strictly)
provable} over $X,\tau$, i.e.\ propositionally entailed by clauses (a
clause) $\psi\sigma$ where $\phi/\psi\in\RulesC$
(Definition~\ref{def:provable}) and~$\sigma$ is a
$\Prop(V)$-substitution (a $V$-substitution) such that
$X,\tau\models\phi\sigma$.
\end{definition}
\noindent
Strict one-step completeness is one of crucial notions in this work. Its
distinctive feature is that strict provability largely dispenses with
propositional reasoning by restricting instantiations to propositional
variables, and by replacing general propositional entailment by the
rather trivial concept of propositional entailment between single
clauses (cf.\ Lemma~\ref{lem:clause-entail}). This plays a central role
in the shallow model construction presented in
Section~\ref{sec:shallowmodels}.

\begin{remark}\label{rem:sosc}
It is shown in~\cite{Schroder06} that the set of all one-step sound
rules is always strictly one-step complete and that the proof system
induced by a one-step complete set of rules is \emph{weakly complete},
i.e.\ proves all valid formulas.
\end{remark}

In the further treatment, we need a further technical condition.
\begin{definition}
A one-step rule $\phi/\psi$ over~$V$ is \emph{injective} if every
variable in~$V$ occurs at most once in~$\psi$. 
\end{definition}
\begin{assumption}\label{ass:injrules}
We assume for the remainder of the paper that \emph{the given rules in
$\Rules$ are injective}. This restriction will be satisfied by the
naturally arising rule sets in our examples; it can always be forced by
introducing new propositional variables and adding premises stating the
equivalence to the original variables (e.g.\ a rule $\top/(\Box a
\modimpl \Diamond a)$ can be replaced by $(a\modiff b)/(\Box a \modimpl
\Diamond b)$).
\end{assumption}


\eat{
\begin{proposition}
A one-step rule $\psi/\phi$ over a set~$V$ of propositional variables is
one-step sound iff $\phi\sigma$ holds in the set $X$ of all valuations
$\tau$ for $V$ such that $\models\psi\tau$, where~$\sigma$ is the
$\powerset(X)$-valuation taking $a\in V$ to the set
\begin{equation*}
\{\tau\in X\mid \tau(a)=\top\}.
\end{equation*}
\end{proposition}
\begin{proof}
We have to show that $\phi\eta$ holds in~$Y$ for every
$\powerset(Y)$-valuation~$\eta$ such that $Y\models\psi\eta$. This
follows by naturality of predicate liftings, applied to the map
$f:Y\to X$, $y\mapsto \lambda a\in V.\, (y\in\eta(a))$.
\end{proof}
}
\noindent
Strictly one-step complete sets of rules are generally more complicated
than one-step complete sets of rules or
axioms~\cite{Pattinson03,Schroder06}. In our terminology, part of the
effort of~\cite{Vardi89} and~\cite{Pauly02} is devoted to finding
strictly one-step complete sets of rules. We now develop a systematic
procedure for turning one-step complete rule sets into strictly one-step
complete ones. For the following, recall that given clauses~$\phi$ and
$\psi$ containing literals~$a$ and~$\neg a$, respectively, a
\emph{resolvent} of~$\phi$ and~$\psi$ (\emph{at~$a$}) is obtained by
removing~$a$ and~$\neg a$ from the clause $\phi\disj\psi$. A set~$\Phi$
of clauses is called \emph{resolution closed} if, for
$\phi,\psi\in\Phi$, all resolvents of~$\phi$ and~$\psi$ are
propositionally entailed by some clause in~$\Phi$.  This is generalised
to rules as follows:
\begin{definition}\label{def:resolution}
A set~$\Rules$ of one-step rules is \emph{resolution closed} if it
satisfies the following requirement. Let $R_1,R_2\in\Rules$, where
$R_1=\phi_1/\psi_1$ and $R_2=\phi_2/\psi_2$. We can assume that~$R_1$
and~$R_2$ have disjoint sets $V_1,V_2$ of propositional variables. Let
$La$ be in~$\psi_1$, and let~$\neg Lb$ be in $\psi_2$ for some
$L\in\ModSig$, so that we have a resolvent $\bar\psi$ of~$\psi_1$ and
$\psi_2[a/b]$ at~$La$; by Assumption~\ref{ass:injrules}, $\bar\psi$ is a
clause over $\Up(V)$ where $V=V_1\cup V_2-\{a,b\}$. Then~$\RulesC$ is
required to contain a rule $R=\phi/\psi$ such
$\phi_1\conj\phi_2[a/b]\PLentails\phi\sigma$ and
$\psi\sigma\PLentails\bar\psi$ for some $V$-substitution $\sigma$; in
this case,~$R$ is called a \emph{resolvent} of~$R_1$ and~$R_2$.
\end{definition}
Resolution closure will play a central role in the following
development, as it forms the syntactic counterpart of strict one-step
completeness.
\begin{remark}
One can construct resolution closed sets by iterated addition of missing
resolvents. Here, an obvious choice for a resolvent $\phi/\psi$ of
$\phi_1/\psi_1$ and $\phi_2/\psi_2$ as above is to take~$\psi$ as the
resolvent $\bar\psi$ of $\psi_1$ and $\psi_2$, and~$\phi$ as
$\phi_1\conj\phi_2[a/b]$, with~$a$ eliminated according to
Remark~\ref{rem:elimination} as~$a$ is not contained in $\psi$ by
Assumption~\ref{ass:injrules}. It is clear that
$\phi_1\conj\phi_2[a/b]/\bar\psi$ is one-step sound if~$R_1$ and~$R_2$
are one-step sound.
\end{remark}
\begin{remark}
Note that our approach is different to 
existing
resolution-based approaches to decision procedures for modal
logic~(e.g.~\cite{DeNivelleEA00}), which rely on translating modal
logic into first-order logic.
\end{remark}
\begin{lemma}\label{lem:resolution}
Let $\psi\in\Clause(V)$, and let $\emptyset\neq\Phi\subseteq\Clause(V)$
be resolution closed. Then $\Phi\PLentails\psi$ iff
$\phi\PLentails\psi$ for some $\phi\in\Phi$.
\end{lemma}


\begin{proof}
The `if' direction is clear. `Only if': W.l.o.g.\ $\psi$ is not a
tautology. We can assume that~$V$ is finite and then prove the
contraposition of the claim by induction over the size of~$V$. Thus
assume, recalling Lemma~\ref{lem:clause-entail}, that~$\Phi$ does not
contain a subclause of~$\psi$.  Pick a clause $\chi\in\Phi$ that
contains a minimal number of literals not in~$\psi$ (this number is
non-zero); w.l.o.g.\ $\chi$ contains a positive literal $a$ such
that~$a$ is not in~$\psi$. Remove all clauses containing~$a$
from~$\Phi$, and remove~$\neg a$ from the remaining clauses and from
$\psi$, obtaining a new set~$\Phi'$ of clauses and a new clause~$\psi'$,
respectively.  Then~$\Phi'$ is resolution closed and does not contain a
subclause of~$\psi'$ (otherwise there exists a clause $\rho\in\Phi$
whose only literal not in~$\psi$ is~$\neg a$, and resolving~$\rho$ with
$\chi$ yields a clause in~$\Phi$ with less literals not in~$\psi$ than
$\chi$, contradiction). By induction we thus have a valuation~$\tau'$
for $V-\{a\}$ satisfying~$\Phi'$ but not~$\psi'$. We extend~$\tau'$ to a
valuation~$\tau$ for~$V$ by putting $\tau(a)=\top$; then~$\tau$
satisfies~$\Phi$ but not~$\psi$.
\end{proof}

\begin{lemma}\label{lem:cong-res}
 $\Rules$ is resolution closed iff~$\RulesC$ is resolution closed.
\end{lemma}
\begin{proof}
The `if' direction is trivial. The `only if' direction follows from the
fact that every rule~$R$ is a resolvent of~$R$ and any congruence rule,
since rules are injective (Assumption~\ref{ass:injrules}).
\end{proof}
\begin{theorem}\label{thm:resolution}
Let~$\Rules$ be one-step complete. Then~$\Rules$ is strictly one-step
complete iff~$\Rules$ is resolution closed.
\end{theorem}
\begin{proof}
\emph{`If':} Let~$X$ be a set, let~$\tau$ be a $\powerset(X)$-valuation,
and let $\chi\in\Clause(\Up(V))$ such that $TX,\tau\models\chi$;
w.l.o.g.\ $\chi$ is not a tautology. By one-step completeness,~$\chi$ is
propositionally entailed by the (non-empty) set of clauses
\begin{equation*}
\Psi=\{\psi\sigma\mid \phi/\psi\in\RulesC,\sigma\textrm{ a
  $\Prop(V)$-substitution}, X,\tau\models\phi\sigma\}.
\end{equation*}
\emph{The set~$\Psi$ is resolution closed:} for $i=1,2$, let
$\phi_i/\psi_i\in\RulesC$ be a rule over~$W_i$ (with~$W_1$,~$W_2$
disjoint), let~$\sigma_i$ be a $\Prop(V)$-substitution such that
$X,\tau\models\phi_i\sigma_i$, and let $\psi_1\sigma_1$ and
$\psi_2\sigma_2$ contain literals~$L\rho$ and
$\neg L\rho$, respectively. Thus,~$\psi_1$ and~$\psi_2$
contain literals~$La$ and~$\neg Lb$,
respectively, where $\sigma_1(a)=\sigma_2(b)=\rho$; let~$\bar\psi$ be
the resolvent of $\psi_1,\psi_2[a/b]$ at~$La$, a clause
over $W=W_1\cup W_2-\{a,b\}$. Then the resolvent of
$\psi_1\sigma_1,\psi_2\sigma_2$ at~$L\rho$ is
$\bar\psi\sigma$, where~$\sigma$ acts like~$\sigma_1$ on $W_1-\{a\}$ and
like~$\sigma_2$ on $W_2-\{b\}$. By resolution closedness of~$\RulesC$
(Lemma~\ref{lem:cong-res}), we have $\phi/\psi\in\RulesC$ and a
$W$-substitution~$\theta$ such that
$\phi_1\conj\phi_2[a/b]\PLentails\phi\theta$ and
$\psi\theta\PLentails\bar\psi$.  Then $X,\tau\models\phi\theta\sigma$, so
that $\psi\theta\sigma\in\Psi$, and
$\psi\theta\sigma\PLentails\bar\psi\sigma$ as required.

By Lemma~\ref{lem:resolution}, it now follows that
$\psi\sigma\PLentails\chi$ for some clause~$\psi\sigma$ in~$\Psi$, where
by Lemma~\ref{lem:clause-entail} necessarily $\sigma(v)\in V$ for every
variable~$v$ in~$\psi$.

\emph{`Only if':} Let $\phi_1/\psi_2,\phi_2/\psi_2\in\Rules$ be rules
over disjoint sets $V_1,V_2$ of variables, where~$\psi_1$ contains
$La$ and~$\psi_2$ contains~$\neg Lb$. Let
$\bar\psi$ denote the resolvent of $\psi_1,\psi_2[a/b]$ at
$La$, a clause over $V=V_1\cup V_2-\{a,b\}$.  Let~$X$ be
the set of satisfying valuations for $\phi_1\land\phi_2[a/b]$, and
define the $\powerset(X)$-valuation~$\tau$ by $\tau(a)=\{\kappa\in X\mid
\kappa(a)=\top\}$. Then $X,\tau\models\phi_1\land\phi_2[a/b]$ and hence
$TX,\tau\models\bar\psi$ by one-step soundness of~$\Rules$. By strict
one-step completeness, it follows that there exists a rule
$\phi/\psi\in\RulesC$ and a $V$-substitution~$\sigma$ such that
$X,\tau\models\phi\sigma$ and $\psi\sigma\PLentails\bar\psi$. By
construction of $X,\tau$, we may conclude from $X,\tau\models\phi\sigma$
that $\phi_1\conj\phi_2[a/b]\PLentails\phi\sigma$ as required.
\end{proof}
\noindent
In summary, strictly one-step complete rule sets can be constructed by
resolving the rules of a one-step complete axiomatisation against each
other. Below, we give examples of strictly one-step complete systems
obtained in this way. In order to simplify the presentation for the case
of graded modal logic and probabilistic modal logic, we use the
following notation.
If~$\phi_i$ is a formula, $r_i \in \mathbb{Z}$ for all $i \in I$, and
$k \in \mathbb{Z}$,
we abbreviate
\begin{equation*}
  \sum_{i\in I} r_i \phi_i  \ge k \equiv \Land_{\substack{J
  \subseteq I\\[2pt]r(J)<k}}
  \quad
  \Big( \Land_{j \in J} \phi_j \modimpl \Lor_{j \notin J}
\phi_j\Big),
\end{equation*}
where $r(J) = \sum_{j \in J} r_j$.  
The formula $\sum_{i\in I} r_i {a_i} \ge k$ translates into
the arithmetic of characteristic functions as suggested by the notation:
\begin{lemma} \label{lemma:sum}
An element $x\in X$ belongs to the interpretation of $\sum_{i\in I} r_i
{a_i} \ge k$ under a $\powerset(X)$-valuation~$\sigma$ iff
\begin{equation*}
  \sum_{i\in I} r_i \cf_{\sigma(a_i)}(x) \ge k,
\end{equation*}
where $\cf_A: X \to \lbrace 0, 1 \rbrace$ is the characteristic
function of $A \subseteq X$.
\end{lemma}
\begin{proof} 
The element~$x$ satisfies the negation of $\sum_{i\in I} r_i {a_i} \ge
k$ iff $r(J)<k$ for $J=\{i\in I\mid x\in\sigma(a_i)\}$ iff $\sum_{i\in I}
r_i \cf_{\sigma(a_i)}(x) < k$.
\end{proof}
We allow ourselves obvious variations of this notation, e.g. $\sum
a_i\le\sum b_j$ in place of $\sum b_j-\sum a_i\ge 0$.

In all the logics of Example~\ref{expl:coalgml}, the resolution process,
applied to known one-step complete rule sets, can be kept under control;
by Theorem~\ref{thm:resolution}, the resulting rule sets are strictly
one-step complete.

\begin{example}\label{expl:resolution}
\begin{longenum}
\item\label{expl:neighbourhood-osc} \emph{Modal logic~$E$:} The empty set
  of rules is one-step complete for neighbourhood frame semantics
  (Example~\ref{expl:coalgml}.\ref{expl:frame-functor}). This set is
  trivially resolution closed.
\item\label{expl:monotone-osc} \emph{Modal logic~$M$:} The one-step rule 
  \begin{equation*}
    \lrule{M}{a\modimpl b}{\Box a\modimpl\Box b}
  \end{equation*}
  is one-step complete for monotone neighbourhood frame semantics
  (Example~\ref{expl:coalgml}.\ref{expl:frame-functor}), and clearly
  resolution closed.
\item\label{expl:k-osc} \emph{Modal logic~$K$:}
The one-step rules
\begin{equation*}
\infrule{a}{\Box a}\qquad
\infrule{a\conj b \modimpl c}{\Box a\conj\Box b \modimpl \Box c}
\end{equation*}
are one-step complete for unrestricted Kripke semantics
(Example~\ref{expl:coalgml}.\ref{expl:power-functor}), i.e.\ for the
modal logic~$K$~\cite{Pattinson03}. The resolution closure~$\Rules$ of
these rules consists of the rules
\begin{equation*}
\infrule{\bigwedge_{i=1}^n a_i \modimpl b}
{\bigwedge_{i=1}^n \Box a_i \modimpl \Box b}
\end{equation*}
for all $n\in\Nat$ (here, strict one-step completeness is also easily
seen directly). Note the similarity between this rule and a
corresponding rule appearing in standard cut-free sequent calculi for
$K$~\cite{TroelstraSchwichtenberg96}; the precise connection between
resolution closure and cut elimination is the subject of further
investigation.
\item \label{expl:kd-osc} \emph{Modal logic~$KD$:} The axiomatisation of
  $K$ is extended to a one-step complete axiomatisation of~$KD$
  (Example~\ref{expl:coalgml}.\ref{expl:kd}) by adding the rule $\neg
  a/\neg\Box a$. Closing the new rule set under resolution leads to the
  rules
  \begin{equation*}
    \infrule{\bigwedge_{i=1}^n a_i \modimpl b}
	    {\bigwedge_{i=1}^n \Box a_i \modimpl \Box b}
	    \quad\textrm{and}\quad
    \infrule{\neg\bigwedge_{i=1}^n a_i}
	    {\neg\bigwedge_{i=1}^n \Box a_i}
  \end{equation*}
  for all $n\in\Nat$ (i.e.\ where the rules of~$K$ apply only to
  positive Horn clauses, the rules of~$KD$ apply to arbitrary Horn
  clauses).
\item\label{expl:cml-osc} \emph{Coalition logic:} In Lemma~6.1
  of~\cite{Pauly02}, the following set of one-step rules for coalition
  logic (Example~\ref{expl:coalgml}.\ref{expl:coalitionlogic}), numbered
  as in loc.\ cit., is implicit:
  \begin{equation*}
    \lrule{1}{\bigvee_{i=1}^n\neg a_i}
	  {\bigvee_{i=1}^n \neg\plbox{C_i} a_i}
    \quad
    \lrule{2}{a}{\plbox{C}a}
    \quad
    \lrule{3}{a \vee b}{\plbox{0}a \vee \plbox{N} b}
  \end{equation*}
  \begin{equation*}
    \lrule{4}{\bigwedge_{i=1}^n a_i \modimpl b}
	  {\bigwedge_{i=1}^n \plbox{C_i} a_i
	    \modimpl\plbox{\bigcup C_i}b}
  \end{equation*}
  where $n\ge 0$, and rules (1) and (4) are subject to the side
  condition that the~$C_i$ are pairwise disjoint. 

  As shown in~\cite{Pauly02}, an axiomatization subsumed by rules
  (1)--(4) is complete for a language including propositional symbols;
  one-step completeness follows by Proposition~\ref{prop:completeness}
  below. The rules are moreover `nearly' resolution closed (full
  resolution closure is not needed in~\cite{Pauly02} due to the use of a
  taylored notion of closed rule set). Resolving rule (4) with rules (2)
  and (3), one obtains the rule schema
  \begin{equation*}
    \lrule{4'}{\bigwedge_{i=1}^n a_i \modimpl b\vee\bigvee_{j=1}^m c_j}
	  {\bigwedge_{i=1}^n \plbox{C_i} a_i
	  \modimpl\plbox{D}b\vee \bigvee_{j=1}^m
	    \plbox{N}c_j}
  \end{equation*}
  where $m,n\ge 0$, subject to the side condition that the~$C_i$ are
  pairwise disjoint subsets of~$D$; this 
  subsumes rules (2)--(4) above. 

  \emph{Resolution closedness} of rules (1) and (4'):
  We discuss
  only the case of resolving $(4')$ against itself; the other case is
  similar. Let one instance of $(4')$ be denoted as in the rule schema,
  and another instance with all entities primed ($a_i'$ etc.). The two
  instances can be resolved in two essentially different ways. The
  subcase where matching is with~$\plbox{D}b$ is straightforward. Thus
  assume w.l.o.g.\ that matching is via
  $\plbox{N}c_1\equiv\plbox{C'_1}a_1'$. Then by the side conditions,
  $D'=N$ and $C'_i=\emptyset$ for $i=2,\dots,n'$. Thus, the resolvent
  has the conclusion
  \begin{equation*}\textstyle
    \bigwedge_{i=1}^n \plbox{C_i} a_i\conj\Land_{i=2}^{n'}\plbox{C'_i} a'_i 
    \modimpl
    \plbox{D}b\disj\plbox{N}b'\disj
	  \bigvee_{j=2}^m \plbox{N}c_j\disj\bigvee_{j=1}^{m'} \plbox{N}c'_j,
  \end{equation*}
  which fits the format of the rule scheme $(4')$. It is easy to check
  that the combined premises imply the required premise for the resolved
  conclusion, and similarly for the side conditions.

\item\label{expl:gml-osc} \emph{Graded modal logic:} The standard
  axiomatization of graded modal logic, weakly complete for a language
  with propositional symbols~\cite{Caro88}, has axioms 
  \begin{equation*}
    \begin{axarray}
    $(G1)$ & $\gldiamond{n+1}a\modimpl\gldiamond{n}a$\\
    $(G2)$ & $\glbox{0}(a\modimpl b)\modimpl\gldiamond{n}a\modimpl
    \gldiamond{n}b$\\
    $(G3)$ & $\glbang{0}(a\land b)\modimpl((\glbang{n_1}a\land
    \glbang{n_2}b)\modimpl\glbang{n_1+n_2}(a\lor b))$\\
    $(N)$ & $\glbox{0}\top$
    \end{axarray}
  \end{equation*}
  where $n,n_1,n_2\in\Nat$, used in a proof system including
  propositional reasoning and the congruence rule (so that $(N)$ induces
  the necessition rule for~$\glbox{0}$). Here, $\glbang{n}\phi$
  abbreviates $\gldiamond{n-1}\phi\land\neg\gldiamond{n}\phi$ for $n>0$,
  and $\neg\gldiamond{0}\phi$ for $n=0$. These axioms may be derived
  from the system of one-step rules
  \begin{gather*}
    \lrule{RG1}{a\modimpl b}{\gldiamond{n+1}a\modimpl\gldiamond{n}b}
    \qquad
    \lrule{A1}
	  {c\modimpl a\lor b}
	  {\gldiamond{n_1+n_2}c\modimpl\gldiamond{n_1}a\lor\gldiamond{n_2}b}\\
    \lrule{A2}
	  {\begin{array}{c}
	a\lor b\modimpl c\\
	a\land b\modimpl d
    \end{array}}
	  {\gldiamond{n_1}a\land\gldiamond{n_2}b\modimpl
	    \gldiamond{n_1+n_2+1}c\lor\gldiamond{0}d}
	  \qquad
    \lrule{RN}{\neg a}{\neg\gldiamond{0}a}
  \end{gather*}
  ($(G1)$ and $(N)$ are easily derived
  from $(RG1)$ and $(RN)$, respectively; $(G2)$ follows by $(A1)$ taking
  $n_2=0$; and $(G3)$ may be derived using $(A1)$ and $(A2)$).
All these rules are subsumed by the rule schema
  \begin{equation*}
   \lrule{G}{\sum_{i=1}^n a_i \le \sum_{j=1}^m b_j}
    {\Land_{i=1}^n \gldiamond{k_i}a_i\modimpl\Lor_{j=1}^m \gldiamond{l_j} b_j},
  \end{equation*}
where $n,m\ge 0$, subject to the side condition $\sum_{i = 1}^n (k_i +
1) \ge 1+ \sum_{j = 1}^m l_j$ (which entails that~$n$ and~$m$ cannot
both be~$0$). One-step soundness of $(G)$ follows from one-step
soundness of the rule system for majority logic proved in the next
example. By the preceding considerations, $(G)$ is weakly complete, and
hence one-step complete by Proposition~\ref{prop:completeness}.

\emph{Resolution closedness of $(G)$:} Take two instances of $(G)$, one
denoted like in the general form of the rule and one with all entities
primed ($a_i'$ etc.), with the resolution taking place w.l.o.g.\ by
matching $\gldiamond{k'_1}a'_1\equiv\gldiamond{l_1}b_1$. The conclusion
of the arising resolvent is
\begin{equation*}
  \Land_{i=1}^n\gldiamond{k_i}a_i\land\Land_{i=2}^{m'}\gldiamond{k'_i}a'_i
  \modimpl\Lor_{i=2}^m \gldiamond{l_j}
  b_j\lor\Lor_{j=1}^{m'}\gldiamond{l'_j}b'_j.
  \end{equation*}
Since $a'_1\equiv b_1$, the premises
\begin{math}
  \sum_{i=1}^n a_i \le \sum_{j=1}^m b_j
\end{math}
and
\begin{math}
  \sum_{i=1}^{n'} a'_i \le \sum_{j=1}^{m'} b'_j
\end{math}
imply
\begin{equation*}
  \sum_{i=1}^n a_i+\sum_{i=2}^{n'} a'_i
\le \sum_{j=2}^m b_j+\sum_{j=1}^{m'} b'_j,
\end{equation*}
and since $k_1=l_1'$, the side conditions
\begin{math}
   \sum_{i = 1}^n (k_i + 1) \ge 1+ \sum_{j = 1}^{m} l_j 
\end{math}
and
\begin{math}
  \sum_{i = 1}^{n'} (k'_i + 1) \ge 1+ \sum_{i = 1}^{m'} l'_j
\end{math}
imply
\begin{equation*}
  \sum_{i = 1}^n (k_i + 1)+\sum_{i = 2}^{n'} (k'_i + 1) \ge 1+
  \sum_{j = 2}^m l_j +  \sum_{j = 1}^{m'} l'_j,
\end{equation*}
so that we arrive again at an instance of $(G)$.
\item \label{expl:majority-osc}\emph{Majority logic:}
  in~\cite{PacuitSalame04}, the extension of the axiomatization of
  graded modal logic with the axioms
  \begin{equation*}
    \begin{axarray}
       $(M1)$ & $Ma\land Mb \modimpl \gldiamond{0}(a\land b)$\\
      $(M2)$ & $Ma\land\glbox{0}(a\to b)\modimpl Mb$\\
      $(M3)$ & $Wa\land Wb\land\gldiamond{n}(\neg a\land\neg b)\modimpl
      \gldiamond{n}(a\land b)$\\
      $(M4)$ & $Wa\land Mb\land\gldiamond{n}(\neg a\land\neg b)\modimpl
      \gldiamond{n+1}(a\land b)$
    \end{axarray}
  \end{equation*}
  is proved to be weakly complete for majority logic including
  propositional symbols. These axioms are derivable from the set of
  rules
  \begin{equation*}
    \lrule{RM1}{a\lor b}{Wa\lor Wb}
    \qquad
    \lrule{RM2}{a\modimpl b\lor c}{Wa\modimpl Wb\lor \gldiamond{0}c}
  \end{equation*}
  \begin{equation*}
    \lrule{RM3}{\begin{array}{c}
	\neg(a\land c)\\
	\neg(b\land c)\\
	a\land b \modimpl d
	\end{array}}
    {Wa \land Wb \land \gldiamond{n}c \modimpl \gldiamond{n}d}
    \qquad
    \lrule{RM4}{\begin{array}{c}
	\neg{a\land b}\\
	a\modimpl c\lor d\\
	b\modimpl c
	\end{array}}
    {Wa\land \gldiamond{n}b\modimpl Wc \lor \gldiamond{n+1}d}
  \end{equation*}
  ($(M2)$, $(M3)$ and $(M4)$ follow directly from $(RM2)$, $(RM3)$ and
  $(RM4)$, respectively; $(RM1)$ proves $Ma\modimpl Wa$, whence $(M1)$
  is obtained from $(RM2)$). These rules and rule $(G)$ for GML are
  subsumed by the rule schema
  \begin{equation*}
    \lrule{M_u}{\textstyle \sum_{i=1}^n a_i + \sum_{r=1}^v c_r + u \le
      \sum_{j=1}^m b_j + \sum_{s=1}^w d_s} 
	  { \Land_{i=1}^n \gldiamond{k_i} a_i \land
	    \Land_{r=1}^v Wc_r \modimpl \Lor_{j=1}^m \gldiamond{l_j} b_j 
	    \lor \Lor_{s=1}^w Wd_s
	  }\;(u\in\mathbb{Z})
  \end{equation*}
  with side conditions $\sum_{i=1}^n (k_i+1) - \sum_{j=1}^m l_j - 1 + w
  - \max (u,0) \ge 0$ and $v - w + 2u \ge 0$ (take $u=1$ for $(RM1)$,
  $u=0$ for $(RM2)$, $(RM4)$, and $(G$), and $u=-1$ for
  $(RM3)$). Resolution closedness is checked analogously as for graded
  modal logic, covering the two cases of resolution at literals
  $\gldiamond{n}a$ and~$Wa$, respectively; in both cases, an instance of
  $M_{u_1+u_2}$ can be taken as a resolvent of an instance of $M_{u_1}$
  and an instance of~$M_{u_2}$.

  \emph{One-step soundness of~$(M_u)$:} Let~$\tau$ be a
  $\powerset(X)$-valuation such that $X,\tau\models \sum_{i=1}^n a_i +
  \sum_{r=1}^v c_r + u \le \sum_{j=1}^m b_j + \sum_{s=1}^w d_s$. Let
  $B\in\Bag(X)$. Using Lemma~\ref{lemma:sum}, we obtain by summation
  over $x\in X$
  \begin{equation*}
    \sum_{i=1}^n B(\sigma(a_i))+\sum_{r=1}^v B(\sigma(c_r))+uB(X)\le
    \sum_{j=1}^m B(\sigma(b_j))+\sum_{s=1}^w B(\sigma(d_s)).
  \end{equation*}
  Now put $p=\lceil B(X)/2\rceil$ (with $\lceil x \rceil = \min
  \lbrace z \in \mathbb{Z} \mid z \geq x
  \rbrace$) so that~$B$ satisfies~$Wa$ iff
  $B(\tau(a))\ge p$. To establish that~$B$ is in the interpretation of
  the conclusion of~$M_u$, it suffices to prove
  \begin{equation*}
    \sum_{i=1}^n(k_i+1) + vp + uB(X) \ge \sum_{j=1}^m l_j + w(p-1) +1.
  \end{equation*}
  By the side conditions, this inequality is equivalent to
  \begin{equation*}
    -2up + uB(X) + \max(u,0)\ge 0,
  \end{equation*}
  which is easily established by distinguishing the cases $B(X)=2p$ and
  $B(X)=2p-1$.
\item\label{expl:pml-osc} \emph{Probabilistic modal logic:} By
  reformulating the one-step complete set of axioms for probabilistic
  modal logic given by~\citeN{CirsteaPattinson07} as one-step rules and
  subsequently applying resolution, one obtains the rules
  \begin{equation*}
    \lrule{P_u}{\sum_{i = 1}^n a_i + u\le\sum_{j=1}^m b_j}
    {\Land_{i=1}^n L_{p_i}a_i\modimpl
      \Lor_{j=1}^m L_{q_j} b_j 
    },
  \end{equation*}
  where $m,n\ge 0$, $m+n\ge 1$, and $u \in \mathbb{Z}$, subject to the
  side condition 
  \begin{align*}
    \textstyle\sum_{i=1}^np_i +     u & \ge  \textstyle\sum_{j = 1}^m q_j
    \textrm{ and}\\
     \textstyle \sum_{i=1}^np_i + u & > 0\quad\textrm{ if }m = 0.
  \end{align*}

  \emph{One-step completeness} of~$(P_u)$:
  The rule schema is one-step complete, as it subsumes the
  following axiomatisation that has been shown to be
  one-step complete in loc.cit.:
   \begin{equation*}
    \laxiom{0}{L_0 a}
    \quad
    \lrule{\top}{a}{L_p a} 
    \quad
    \lrule{>1}{\neg a\disj\neg b}{\neg L_p a\disj\neg L_q b}\;(p+q>1)
  \end{equation*}
  \begin{equation*}
    \lrule{1}{a\disj b}{L_p a\disj L_q b}\;(p+q=1)
  \end{equation*}
  \begin{equation*}
    \lrule{\cf}{\sum_{i=1}^r c_i = \sum_{j=1}^s \bar d_j} {\Land_{i=1}^r
    L_{u_i}c_i\conj \Land_{j=2}^s L_{(1-v_j)}d_j \modimpl L_{v_1}d_1},
  \end{equation*}
  where $\bar d_1=d_1$ and $\bar d_j=\neg d_j$ for $j\ge 2$, and rule
  $(\cf)$ is subject to the side condition
  \begin{equation*}
    \sum_{j=1}^s v_j = \sum_{i=1}^r u_i.
  \end{equation*}
  These rules are subsumed by the rule schema $(P_u)$, as
  follows. Rule $(0)$: take $m=1$, $n=0$, $u=0$, $q_1=0$. Rule $(\top)$:
  take $m=1$, $n=0$, $u=1$. Rule $>1$: take $n=2$, $m=0$, $u=-1$. Rule
  $(1)$: take $n=0$, $m=2$, $u=-1$. Rule $(\cf)$: take $m=1$,
  $n=r+s-1$, $u=1-s$, and instantiate~$b_i$ to~$c_i$ for
  $i=1,\dots,r$, $b_i$ to~$d_{i-r+1}$ for $i=r+1,\dots,r+s-1$,~$a_1$ to
  $d_1$,~$q_i$ to~$u_i$ for $i=1,\dots,r$,~$q_i$ to $1-v_{i-r+1}$ for
  $i=r+1,\dots,r+s-1$, and~$p_1$ to~$v_1$.

  \emph{One-step soundness}: Analogously to the previous
  example, using additionally that one always has $P(X)=1$. 
  
  \emph{Resolution closedness:} Analogously as for graded modal logic;
  as a resolvent of an instance of~$P_{u_1}$ and an instance of
  $P_{u_2}$, one can take an instance of $P_{u_1+u_2}$.
\end{longenum}
  \end{example}

\section{The Shallow Model Construction}\label{sec:shallowmodels}

We now present the announced generic shallow model construction, which
is based on strictly one-step complete axiomatisations. The construction
generalises results from~\cite{Vardi89} (where the use of
axiomatisations is implicit in certain lemmas).

\begin{definition}\label{def:demand}
The set~$\ModAts(\phi)$ of (top level) \emph{modal atoms} of a formula
$\phi$ is defined recursively by
$\ModAts(\phi\conj\psi)=\ModAts(\phi)\cup\ModAts(\psi)$,
$\ModAts(\neg\phi)=\ModAts(\phi)$, and $\ModAts(L\rho)=\{L\rho\}$. (Note
that $\phi\in\Prop(\ModAts(\phi))$.) A \emph{pseudovaluation} is a
conjunctive clause~$H$ over $\Up(\FLang(\ModSig))$, represented as a set
of literals (i.e.\ pseudovaluations are identified modulo contraction
and reordering of literals, which does not affect the set $\ModAts(H)$
of modal atoms). A pseudovaluation is \emph{consistent} if it is
consistent as an $\FLang(\ModSig)$-formula. 
We say that~$H$ is a \emph{pseudovaluation for~$\phi$}
if $\ModAts(H)\subseteq\ModAts(\phi)$ and $H\PLentails\phi$. If
$\phi/\psi$ is a rule in $\RulesC$ and~$\sigma$ is a substitution such
that $\psi\sigma\in\Clause(\ModAts(H))$ and $H\PLentails\neg\psi\sigma$,
then the negated instance $\neg\phi\sigma$ of the premise~$\phi$ is a
\emph{demand} of~$H$.
\end{definition}
This generalises the notion of demand \cite[Definition
  6.43]{BlackburnEA01} to a coalgebraic setting.  Note that by the dual
of Lemma~\ref{lem:clause-entail}, all demands of a pseudovaluation~$H$
are contained in~$H$ when regarded as sets of literals, unless $H$ is
propositionally inconstent (i.e.\ contains both~$L\rho$ and $\neg L\rho$
for some modal atom~$L\rho$).
\begin{lemma}\label{lem:pseudoval}
Every consistent formula has a consistent pseudovaluation.
\end{lemma}
\begin{proof}
If~$\phi$ is consistent, then one of the conjunctive clauses from its
disjunctive normal form (DNF) is consistent and hence is a consistent
pseudovaluation for~$\phi$.
\end{proof}

\begin{lemma}\label{lem:demands}
Every demand of a consistent pseudovaluation is consistent.
\end{lemma}
\begin{proof}
By contraposition: Let~$H$ be a pseudovaluation, and let $\phi/\psi$ be
a rule in~$\RulesC$ such that $\psi\sigma\in\Clause(\ModAts(H))$ and
$H\PLentails\neg\psi\sigma$.  If the demand $\neg\phi\sigma$ is
inconsistent, then $\phi\sigma$ is provable; hence, $\psi\sigma$ is
provable using $\phi/\psi$, and consequently~$H$ is inconsistent.
\end{proof}

\begin{definition}
A \emph{supporting Kripke frame} of a $T$-coalgebra $(X,\xi)$ is a
Kripke frame $(X,\Kripke)$ (consisting of a set~$X$ and a transition
relation $\Kripke\subseteq X\times X$) such that for each $x\in X$,
\begin{equation*}
\xi(x)\in T\{y\mid x\Kripke y\}\subseteq TX.
\end{equation*}
\end{definition}

\begin{lemdefn}\label{lemdefn:submodel}
If a coalgebra $C=(X,\xi)$ is equipped with a supporting Kripke frame
$(X,\Kripke)$, then for every state $x\in X$, the set~$X_x$ of states
reachable from~$x$ in~$(X,\Kripke)$ is the carrier of a subcoalgebra
$C_x=(X_x,\xi_x)$ of~$C$, the \emph{submodel generated by~$x$}.
\end{lemdefn}
Note that by Proposition~\ref{prop:adequacy}, $y\models_{C_x}\phi$ iff
$y\models_C\phi$ for $y\in X_x$.

\begin{definition}
A \emph{shallow tableau} is a Kripke frame $(X,\Kripke)$ with a distinguished
\emph{root} $H_0\in X$ such that~$X$ is a set of pseudovaluations, every
state is reachable from~$H_0$, for all $H,G\in X$,
\begin{equation*}
H \Kripke G\implies\textrm{~$G$ is a pseudovaluation for a demand of~$H$},
\end{equation*}
and for every demand~$\phi$ of $H\in X$ there exists a pseudovaluation
$G\in X$ for~$\phi$ such that $H \Kripke G$. Given a formula~$\phi$, a
\emph{shallow tableau for~$\phi$} is a shallow tableau whose root is a
pseudovaluation for~$\phi$.

A \emph{shallow tableau model} is a $T$-coalgebra $C=(X,\xi)$ which has
a supporting Kripke frame $(X,\Kripke)$ such that $(X,\Kripke)$ is a shallow tableau
and the \emph{truth lemma} 
\begin{equation*}
H\PLentails\chi\implies H\models_C\chi\quad\textrm{ for all
  $\FLang(\ModSig)$-formulas $\chi$}
\end{equation*}
holds for all $H\in X$ (hence in particular $H\models_C\chi$ if~$H$ is a
pseudovaluation for~$\chi$).
\end{definition}
A shallow tableau is almost a dag, except that in the presence of the
rule $\bot/\bot$ (cf.\ Section~\ref{sec:deduction}) the pseudovaluation
$\top$ is a pseudovaluation for one of its own demands. Explicitly:
\begin{proposition}
A shallow tableau $(X,\Kripke)$ with root~$H_0$ is, up to a possible loop at
the state~$\top$, a dag of depth at most the depth of~$H_0$, and the
branching degree at $H\in X$ is exponentially bounded in~$|H|$.
\end{proposition}
\begin{proof}
The first claim follows from the fact the the depth of all demands of a
pseudovaluation~$H$ is strictly less than the depth of~$H$. To prove the
bound on branching, note that pseudovaluations for demands of~$H$ are
conjunctive clauses over the set of subformulas of~$H$.
\end{proof}
\begin{lemma}\label{lem:tableau}
If a formula~$\phi$ has a pseudovaluation~$H_0$ such that all demands of
$H_0$ are consistent, then there exists a shallow tableau for~$\phi$.
\end{lemma}
(By Lemmas~\ref{lem:pseudoval} and~\ref{lem:demands}, the conditions of
the above lemma hold in particular if~$\phi$ is satisfiable.)
\begin{proof}
Let~$Z$ consist of~$H_0$ and all consistent pseudovaluations, and for
$H,G\in Z$ put $H\bar \Kripke G$ iff~$G$ is a pseudovaluation for a demand of
$H$. Let $(X,\Kripke)$ be the subframe of $(Z,\bar \Kripke)$ generated by~$H_0$
(i.e.~$X$ is the set of states reachable from~$H_0$ in $(Z,\bar \Kripke)$,
and $\Kripke=\bar \Kripke\cap (X\times X)$). By the assumption on~$H_0$ and
Lemmas~\ref{lem:pseudoval} and~\ref{lem:demands}, $(X,\Kripke)$ is a shallow
tableau for~$\phi$.
\end{proof}

\begin{theorem}\label{thm:tableau}
If~$\Rules$ is strictly one-step complete, then every shallow tableau is
a supporting Kripke frame of a shallow tableau model.
\end{theorem}
\begin{proof}
Let $(X,\Kripke)$ be a shallow tableau; we have to construct a shallow tableau
model $C=(X,\xi)$ for which $(X,\Kripke)$ is a supporting Kripke frame.  To
begin, note that to ensure the truth lemma, it suffices that~$C$ is
\emph{coherent} in the sense that for $H\in X$ and $Y=\{G\mid H \Kripke G\}$,
\begin{equation*}
  H\PLentails L\rho\iff\xi(H)\in\Sem{L}_Y\{G\in Y\mid
  G\models_{C_G}\rho\}\textrm{ for all $L\rho\in\ModAts(H)$}
\end{equation*}
(cf.\ Lemma and Definition~\ref{lemdefn:submodel}) : note that $\{G\in
Y\mid G\models_{C_G}\rho\}=\Sem{\rho}_C\cap Y$, so that by naturality of
predicate liftings, coherence implies that
\begin{equation*}
H\PLentails L\rho\iff H\models_CL\rho
\textrm{ for all $L\rho\in\ModAts(H)$}.
\end{equation*}
The extension to propositional consequences of~$H$ is then
straightforward (noting that for $L\rho\in\ModAts(H)$,
either $H\PLentails L\rho$ or
$H\PLentails\neg L\rho$).

We construct a coherent coalgebra structure~$\xi$ by induction over the
depth of pseudovaluations. Thus, let $H\in X$, put $Y=\{G\mid H \Kripke G\}$,
and assume that~$\xi$ is already constructed for all pseudovaluations of
smaller depth in~$X$, in particular for all states~$G$ reachable from~$H$
in $(X,\Kripke)$. Thus, the submodel~$C_G$ generated by such a state~$G$ is
already defined, and coherence at~$G$ is unaffected by the construction
of~$\xi(H)$.

We have to prove that there exists $\xi(H)\in TY\subseteq TX$ satisfying
the coherence condition. Assume the contrary. Let~$V$ be the set of
propositional variables~$b_\rho$, where
$L\rho\in\ModAts(H)$ for some~$L$. Let
$\theta\in\Clause(\Up(V))$ consist of the literals
$\neg Lb_\rho$ for $L\rho\in H$ and
$Lb_\rho$ for $\neg L\rho\in H$.  By
assumption, $TY,\tau^Y\models \theta$, where~$\tau^Y$ is the
$\powerset(Y)$-valuation taking~$b_\rho$ to $\{G\in Y\mid
G\models_{C_G}\rho\}$. By strict one-step completeness, it follows that
$\psi\eta\PLentails\theta$ for a rule $\phi/\psi$ in~$\RulesC$ and a
$V$-substitution~$\eta$ such that $Y,\tau^Y\models\phi\eta$.  By
construction of~$\theta$, $H\PLentails\neg\theta\sigma$ and hence
$H\PLentails\neg\psi\eta\sigma$. Thus, $\neg\phi\eta\sigma$ is a demand
for~$H$, and hence there exists in~$Y$ a pseudovaluation~$G$ for
$\neg\phi\eta\sigma$. By the truth lemma for~$G$,
$G\models_{C_G}\neg\phi\eta\sigma$, in contradiction to
$Y,\tau^Y\models\phi\eta$.
\end{proof}

\begin{corollary}\label{cor:tableau}
If~$\Rules$ is strictly one-step complete, then the following are
equivalent for an $\FLang(\ModSig)$-formula~$\phi$.
\begin{enumerate}
\item\label{item:sat}~$\phi$ is satisfiable.
\item\label{item:cons}~$\phi$ is consistent.
\item\label{item:pseudovalcons}~$\phi$ has a pseudovaluation~$H$ such that
  all demands of~$H$ are consistent.
\item\label{item:pseudovalsat}~$\phi$ has a pseudovaluation~$H$ such that
  all demands of~$H$ are satisfiable.
\item \label{item:tableau} There exists a shallow tableau for~$\phi$.
\item \label{item:tableau-model}~$\phi$ is satisfiable at the root of a
shallow tableau model.
\end{enumerate}
\end{corollary}
\begin{proof}
\emph{(\ref{item:sat})$\implies$(\ref{item:cons}):} By soundness.

\emph{(\ref{item:cons})$\implies$(\ref{item:pseudovalcons}):} By
Lemmas~\ref{lem:pseudoval} and~\ref{lem:demands}

\emph{(\ref{item:pseudovalcons})$\implies$(\ref{item:tableau}):} By
Lemma~\ref{lem:tableau}.

\emph{(\ref{item:tableau})$\implies$(\ref{item:tableau-model}):} By
Theorem~\ref{thm:tableau}.

\emph{(\ref{item:tableau-model})$\implies$(\ref{item:sat}):}
Trivial.

\emph{(\ref{item:pseudovalcons})$\iff$(\ref{item:pseudovalsat}):} By the
equivalence (\ref{item:sat})$\iff$(\ref{item:cons}) already established.
\end{proof}
The above implies in particular that the proof system is \emph{weakly
complete}, i.e.\ proves all valid formulas; this reproves a result
of~\cite{Pattinson03}. By Remark~\ref{rem:sosc}, we obtain moreover that
coalgebraic modal logic has the shallow model property:

\begin{corollary}[Shallow model property]\label{cor:smp}
Every satisfiable $\FLang(\ModSig)$-formula~$\phi$ is satisfiable in a
\emph{shallow model}, i.e.\ in a $T$-coalgebra that has a supporting
Kripke frame $(X,\Kripke)$ which has \emph{final} state~$x_\top$, i.e.\
$x_\top \Kripke x$ implies $x=x_\top$, and which, up to a possible loop at
$x_\top$, is a dag of depth at most the depth of~$\phi$ and of size at
most~$3^n$, where~$n$ is the number of subformulas of~$\phi$.
\end{corollary}
\begin{proof}
All that remains to be checked is the bound on the size: every state in
a shallow tableau is a set representing a conjunctive clause over
subformulas of~$\phi$, in which a given subformula may occur as a
positive literal, as a negative literal, or not at all.
\end{proof}

\section{Shallow Proofs}\label{sec:shallowproofs}

The satisfiability criterion of Corollary~\ref{cor:tableau} can be
rephrased in terms of a \emph{shallow proof property}. This property can
be proved semantically by dualising Corollary~\ref{cor:tableau}, as done
in the proof of Corollary~\ref{cor:shallow-proofs} below. Alternatively,
the shallow proof property can be established purely syntactically,
without any reference to models; we present such an argument in the
proof of Theorem~\ref{thm:shallowproofs} below. The shallow model
construction presented in the previous section is however of independent
interest.

\begin{corollary}[Shallow Proof Property] \label{cor:shallow-proofs}
Let~$\Rules$ be strictly one-step complete. Then an
$\FLang(\ModSig)$-formula $\phi$ is provable iff for each clause~$\rho$
in the conjunctive normal form (CNF) of~$\phi$, there exists a rule
$\chi/\psi\in\RulesC$ and a substitution~$\sigma$ such that
$\psi\sigma\PLentails\rho$ and $\chi\sigma$ is provable.
\end{corollary}
\begin{proof}
The `if' direction is trivial; we prove `only if'. Dualizing the
implication (\ref{item:pseudovalcons})$\implies$(\ref{item:cons}) in
Corollary~\ref{cor:tableau} yields
\begin{quote}
  if~$\phi$ is provable then each pseudovaluation~$H$ for~$\neg\phi$ has a
  demand~$\chi$ such that~$\neg\chi$ is provable.
\end{quote}
Now let~$\rho$ be a clause in the CNF of~$\phi$. Then~$\neg\rho$ is a
conjunctive clause in the DNF of~$\neg\phi$, in particular a
pseudovaluation for~$\neg\phi$. By the above condition, there exists a
rule $\chi/\psi\in\RulesC$ and a substitution~$\sigma$ such that
$\neg\rho\PLentails\neg\psi\sigma$, hence $\psi\sigma\PLentails\rho$,
and~$\chi\sigma$ is provable.
\end{proof}

In a purely syntactic formulation of the shallow proof property, we have
to replace strict completeness by closedness under resolution. The
statement thus takes the following form.


\begin{theorem}[Shallow Proof Property]\label{thm:shallowproofs}
Let~$\Rules$ be resolution closed. Then an $\FLang(\ModSig)$-formula
$\phi$ is provable under~$\Rules$ iff for each clause~$\rho$ in the CNF
of~$\phi$, there exists a rule $\chi/\psi\in\RulesC$ and a substitution
$\sigma$ such that $\psi\sigma\PLentails\rho$ and $\chi\sigma$ is
provable.
\end{theorem}
(This reproves Corollary~\ref{cor:shallow-proofs}, as strict one-step
completeness implies resolution closedness by
Theorem~\ref{thm:resolution}.)
\begin{proof}
Again, `if' is trivial, and we prove `only if'. Let~$\phi$ be provable,
and let~$\rho$ be a clause in the CNF of~$\phi$. Then~$\rho$ is
provable. By definition of the proof system,~$\rho$ is propositionally
entailed by the set of clauses
\begin{equation*}
  \Phi=\{\psi\sigma\mid\chi/\psi\in\RulesC,\chi\sigma\textrm{ provable}\}.
\end{equation*}
One shows analogously as in the `if' direction of the proof of
Theorem~\ref{thm:resolution} that~$\Psi$ is resolution closed. By
Lemma~\ref{lem:resolution}, there exists~$\psi\sigma$ in~$\Phi$ such
that $\psi\sigma\PLentails\rho$.
\end{proof}
We hope that both proofs of the shallow proof property provide a
handle for generalizations to logics outside rank~1.

One application of the shallow proof property is
\begin{proposition}\label{prop:completeness}
Let~$\ModSig$ contain an infinite set $U$ of propositional symbols,
modelled as in Remark~\ref{rem:propsymb} over a
functor~$T_U$ of the form $T_UX=TX\times\powerset(U)$. Then the proof system
induced by~$\Rules$ is weakly complete iff~$\Rules$ is one-step
complete.
\end{proposition}
\begin{proof}
W.l.o.g.~$\Rules$ is resolution closed (one can close under resolution,
thereby affecting neither completeness nor one-step completeness).  The
`if' direction is known (cf.\ Remark~\ref{rem:sosc}). To prove the `only
if' direction, let $\psi\in\Clause(\Up(V))$, let~$X$ be a set, and let
$\tau$ be a $\powerset(X)$-valuation such that
$T_UX,\tau\models\psi$. Since~$U$ is infinite and~$V$ may be assumed to
be finite, we can assume w.l.o.g.\ that $V\subseteq U$. Let~$\phi$
denote the propositional theory of~$\tau$, i.e.\ the conjunction of all
contracted clauses~$\chi$ over~$V$ such that $X,\tau\models\chi$. Then
one checks as in the proof of Theorem~17 in~\cite{Schroder06} that the
rule $\phi/\psi$ is one-step sound. By Lemma~16 in~\cite{Schroder06},
there exists a $\Prop(V)$-substitution~$\sigma$ such that $\phi\sigma$
and $\phi\modimpl(a\modiff\sigma(a))$ (for each~$a\in V$) are
propositional tautologies. Since $V\subseteq U$, we can regard
$\phi\sigma$ as an $\FLang(\ModSig)$-formula. As such, $\phi\sigma$ is
valid. By soundness of $\phi/\psi$, it follows that~$\psi\sigma$, again
regarded as an $\FLang(\ModSig)$-formula, is valid, hence provable by
weak completeness. By the shallow proof property
(Theorem~\ref{thm:shallowproofs}), there exist a rule $\chi/\rho$ over
$W$ and a $\Prop(V)$-substitution~$\theta$ such that $\chi\theta$ is
provable and $\rho\theta\PLentails\psi\sigma$. By
Lemma~\ref{lem:clause-entail} and Assumption~\ref{ass:injrules}, it
follows that there exists a $V$-substitution~$\kappa$ such that
$\sigma(\kappa(b))=\theta(b)$ for all $b\in W$ and
$\rho\kappa\PLentails\psi$.

It remains to prove that $X,\tau\models\chi\kappa$. From
$X,\tau\models\phi$ and the construction of~$\sigma$, we obtain
$X,\tau\models a\modiff\sigma(a)$ for all $a\in V$ and hence
$X,\tau\models\kappa(b)\modiff\theta(b)$ for all $b\in W$, so that the
goal follows from $X,\tau\models\chi\theta$.
\end{proof}

\begin{remark}
  In the above result, the assumption that $\ModSig$ contains enough
  propositional symbols is essential. E.g.\ in cases like coalition logic
  or probabilistic modal logic where the logic collapses into triviality
  without propositional symbols, the empty set of rules is complete, but
  not one-step complete.
\end{remark}

  The proof-theoretic content of Theorem~\ref{thm:shallowproofs} goes
  beyond the mere fact that proofs are shallow. The theorem asserts that
  if the rule system is resolution closed, then propositional reasoning
  can always be limited to decomposing a formula into the clauses of its
  CNF and propositional entailment (i.e.\ by
  Lemma~\ref{lem:clause-entail} essentially containment) between
  clauses. Moreover, shallow proofs witness a \emph{weak subformula
  property}: every provable formula has a proof that mentions only
  propositional combinations of subformulas. Formally:

\begin{theorem}[Weak subformula property]
  Suppose that~$\Rules$ is resolution closed and~$\phi$ is derivable
  under~$\Rules$. Then there exists a proof of~$\phi$ that 
  mentions only propositional combinations of subformulas of~$\phi$.
  \end{theorem}

\begin{proof}
 Assume that~$\phi$ is derivable under~$\Rules$ and~$\rho$ is a clause
 of the CNF of~$\phi$; w.l.o.g.~$\rho$ is not a tautology. By Theorem
 \ref{thm:shallowproofs} we find a rule $\chi / \psi \in \Rules$ and a
 substitution~$\sigma$ such that $\psi \sigma \PLentails \rho$ and $\chi
 \sigma$ is provable under $\Rules$; by
 Lemma~\ref{lem:clause-entail},~$\rho$ contains~$\psi \sigma$, hence we
 can assume w.l.o.g.\ that~$\sigma$ maps propositional variables to
 subformulas of $\rho$. As~$\chi$ is a purely propositional formula, the
 substituted premise $\chi \sigma$ is a propositional combinations of
 subformulas of~$\rho$, hence also of~$\phi$. The claim now follows
 inductively.
\end{proof}

  As a consequence, it is immediate that $\FLang(\ModSig)$ is a conservative
  extension of any sublanguage $\FLang(\ModSig_0)$ induced by 
  a sub-signature $\ModSig_0\subseteq\ModSig$:

  \begin{corollary}[Conservativity]
  Suppose~$\Rules$ is resolution closed, $\Lambda_0 \subseteq \Lambda$
  is a sub-signature and~$\Rules_0$ consists of those $\phi/\psi \in
  \Rules$ that mention only modal operators in $\Lambda_0$. Then a
  formula $\phi \in \FLang(\ModSig_0)$ is $\Rules$-derivable iff it is
  $\Rules_0$-derivable. 

  In particular, if~$\Rules$ is weakly complete for~$\FLang(\ModSig)$,
  then~$\Rules_0$ is weakly complete for~$\FLang(\ModSig_0)$.
  \end{corollary}

  \begin{example}
  From completeness of the rules~$(M_u)$
  for majority logic
  (Example~\ref{expl:resolution}.\ref{expl:majority-osc}), we obtain
  that the  rules
\begin{equation*}
  \lrule{W_u}{\textstyle \sum_{r=1}^v c_r + u \le
      \sum_{s=1}^w d_s} 
	{ \Land_{r=1}^v Wc_r \modimpl \Lor_{s=1}^w Wd_s
  }\;(u\in\mathbb{Z})
\end{equation*}
with side conditions $w - 1 - \max (u,0) \ge 0$ and $v - w + 2u \ge 0$
form a complete axiomatisation of the majority operator~$W$
alone. (\citeN{Pauly05} considers a similar language, but without
nesting of modal operators in formulas.)
\end{example}

  \section{A Generic PSPACE Algorithm}\label{sec:pspace}

\noindent
We will now exploit the shallow model result (Corollary
\ref{cor:tableau}) to design a decision procedure for satisfiability in
the spirit of~\cite{Vardi89}. This requires one more preparatory step:
since resolution closed rule sets are in general infinite, we must
ensure that we never need to instantiate a rule in such a way that the
conclusion contains the same literal twice; otherwise, determining the
demands of a given pseudovaluation (Definition~\ref{def:demand}) might
require checking infinitely many rules. This is formally captured as
follows.

\begin{definition}
An instance $\phi \sigma / \psi \sigma$ of a rule $\phi /\psi$ is
\emph{contracted} if the clause $\psi\sigma$ is contracted
(Definition~\ref{def:propstuff}). In this case, if~$H$ is a
pseudovaluation (Definition~\ref{def:demand}) such that
$\psi\sigma\in\Clause(\ModAts(H))$ and $H\PLentails\neg\psi\sigma$, the
demand $\neg\phi\sigma$ of~$H$ is called an \emph{essential demand}. We
say that a set~$\Rules$ of rules is \emph{closed under contraction} if
for every $V$-instance $\phi \sigma / \psi \sigma$ of a rule $\phi /
\psi$ over $V$ in~$\Rules$, there exists a contracted $V$-instance
$\phi'\sigma' / \psi'\sigma'$ of a rule $\phi' /\psi' \in \Rules$ such
that $\psi'\sigma'$ propositionally entails $\psi\sigma$ and
$\phi\sigma$ propositionally entails $\phi' \sigma'$.
\end{definition}

I.e.\ a rule set is closed under contraction if every instance of a rule
that duplicates literals in the conclusion can be replaced by a
contracted instance of a different rule. Not all the rule sets discussed
in Example \ref{expl:resolution} satisfy this property, but they can
easily be closed under contraction: just add a rule $\phi'/\psi'$ for
every rule $\phi/\psi$ over~$V$ in~$\Rules$ and every $V$-substitution
$\sigma$, where~$\phi'$ is some suitably chosen propositional equivalent
of $\phi\sigma$ and~$\psi'$ is obtained from $\psi\sigma$ by removing
duplicate literals. It is clear that the new rules remain one-step
sound. Note that extending the rule set trivially preserves strict
one-step completeness, so that there is no need to close the extended
rule set under resolution again.

For convenience, we introduce further notation for propositional
formulas: if $r \in \mathbb{Z}-\{0\}$ and~$\phi$ is a formula, then we
put
\begin{equation*}
  \sgn(r) \phi = \begin{cases}
  \phi & r > 0 \\[-0.5ex]
  \neg \phi & r < 0.
  \end{cases}
\end{equation*}

\begin{example} \label{expl:contraction}
  \begin{longenum}
  \item The strictly one-step complete rule sets of
    Examples~\ref{expl:resolution}.\ref{expl:neighbourhood-osc}--\ref{expl:cml-osc}
    ($E$, $M$, $K$, $KD$, and coalition logic) are easily seen to be closed
    under contraction, essentially because in all relevant rule schemas,
    the premise is a clause of the same general format as the
    conclusion.
  \item \emph{Graded modal logic:} The rule schema $(G)$ of
    Example~\ref{expl:resolution}.\ref{expl:gml-osc} fails to be
    closed under contraction, as duplicating literals in the conclusion
    substantially affects both the premise and the side condition. We
    can close $(G)$ under contraction as described above; this results in
    the rule schema
    \begin{equation*}
      \lrule{G'}{\sum_{i=1}^n r_i {a_i}\ge 0}
              {\Lor_{i=1}^n \sgn(r_i)\gldiamond{k_i}a_i},
    \end{equation*}
    where $n\ge 1$ and $r_1,\dots,r_n \in \mathbb{Z}-\{0\}$, subject to
    the side condition $\sum_{r_i < 0} r_i (k_i+1)\ge 1+\sum_{r_i > 0}
    r_i k_i$.
\item\label{expl:majority-contraction} \emph{Majority logic:} Similarly,
  closing the rule schema $(M_m)$ for majority logic under contraction
  yields the rule schema
  \begin{equation*}
    \lrule{M_m'}{m \le \sum_{i=1}^n r_ia_i +
      \sum_{j=1}^v s_jb_j}
	    { \Lor \sgn(r_i)\gldiamond{k_i} a_i \lor \Lor \sgn(s_j)Wb_j  
	    }\;(r_i,s_j\in\mathbb{Z}-\{0\},m\in\mathbb{Z})
  \end{equation*}
  with side conditions $\sum_{r_i < 0} r_i (k_i+1)- (\sum_{r_i > 0}
  r_i k_i) - 1 + \sum_{s_j>0}s_j- \max(m,0)\ge 0$ and 
  $2m-\sum s_j\ge 0$.  
\item\label{expl:pml-contraction} \emph{Probabilistic modal logic:} The
  rule schema $(P_k)$ of
  Example~\ref{expl:resolution}.\ref{expl:pml-osc} fails to be closed
  under contraction. Closure under contraction as described above leads
  to the rule schema
  \begin{equation*}
    \lrule{P'_k}{\sum_{i = 1}^n r_i a_i \ge k}
      {\Lor_{1 \leq i \leq n} \sgn(r_i) L_{p_i} a_i}
  \end{equation*}
  where $n\ge 1$ and $r_1,\dots,r_n \in\mathbb{Z}-\{0\}$, subject to the
  side condition
  \begin{align*}
    &\textstyle\sum_{i = 1}^n r_i p_i \le k,\textrm{ and}\\
    \textrm{if $\forall i.\, r_i < 0$ then}
    &\textstyle\sum_{i = 1}^n r_i p_i < k.
  \end{align*}
\end{longenum}
\end{example}

The crucial property of
contraction closed rule sets is
\begin{lemma}
If~$\Rules$ is closed under contraction, then all the demands of a
pseudovaluation are satisfiable iff all its essential demands are
satisfiable.
\end{lemma}
\begin{proof}
The `only if' direction is trivial. We prove `if': Let~$\Rules$ be
closed under contraction. Then also~$\Rules_C$ is closed under
contraction, since instances of the congruence rule never contain
duplicate literals. Thus, every demand of a pseudovaluation~$H$ is
propositionally entailed by an essential demand.
\end{proof}

Thus we can extend Corollary~\ref{cor:tableau} as follows.
\begin{corollary}\label{cor:tableau-contracted}
If~$\Rules$ is strictly one-step complete and closed under contraction,
then an $\FLang(\ModSig)$-formula~$\phi$ is satisfiable iff~$\phi$ has a
pseudovaluation~$H$ such that all essential demands of~$H$ are satisfiable.
\end{corollary}

In the algorithm suggested by Corollary~\ref{cor:tableau-contracted}, we
will encode demands, which are themselves too large to be passed
around directly, by the rules that induce them. Here, we need to
represent rules by suitable \emph{codes}, i.e.\ strings over some
alphabet, since a naive direct representation of rules would in
particular have to deal with rule premises of potentially exponential
size. 
\begin{definition}
We say that a rule $R\in\Rules$ \emph{matches} a clause
$\rho\equiv\Lor_{i=1}^n\epsilon_iL_i\phi_i$ if the conclusion of~$R$ is
of the form $\Lor_{i=1}^n\epsilon_iL_ia_i$. In this case, let
$\sigma(R,\rho)$ denote the arising substitution
$[\phi_i/a_i]_{i=1,\dots,n}$. Two rules matching the same clause are \emph{equivalent}
if their premises are propositionally equivalent; equivalence classes
$[R]$ are called \emph{$\Rules$-matchings}. The code of~$R$ is also a
\emph{code} for~$[R]$.
\end{definition}
\noindent
We fix some size measures for the representation of formulas and rules:
\begin{definition}\label{def:size}
The size $\size(a)$ of an integer~$a$ is $\lceil \log_2 (|a|+1) \rceil$,
where $\lceil r \rceil = \min \lbrace z \in \mathbb{Z} \mid z \geq r
\rbrace$ as usual. The size $\size(p)$ of a rational number $p=a/b$,
with $a,b$ relatively prime, is $1+\size(a)+\size(b)$. The \emph{size}
$|\phi|$ of a formula~$\phi$ over~$V$ is defined by counting~$1$ for
each propositional variable, boolean operator, or modal operator, and
additionally the size of each index of a modal operator. (In the
examples, indices are either numbers, with sizes as above, or subsets of
$\{1,\dots,n\}$, assumed to be of size~$n$.)
\end{definition}
\begin{assumption}\label{ass:formula-rep}
We assume a reasonable encoding of modal formulas in which boolean
operators take up constant space and modal operators take up space
according to a given coding of~$\ModSig$; we assume that this coding is
in~$\NP$ (i.e.\ it is decidable in~$\NP$ whether a given code is a valid
code for a modal operator in~$\ModSig$). Graded or probabilistic modal
operators are assumed to be coded in binary, with sizes according to
Definition~\ref{def:size}.
\end{assumption}
\begin{example}\label{expl:codes}
For the rules of Examples~\ref{expl:resolution}
and~\ref{expl:contraction}, we just take the parameters of a rule as its
code in the obvious way. E.g.\ the code of an instance of~$(P'_k)$ as
displayed in Example~\ref{expl:contraction}.\ref{expl:pml-contraction}
consists of~$n$,~$k$, the~$r_i$, and the~$p_i$. The size of the code is
determined by the sizes of these numbers plus separating letters, say,
$\sum(1+\size(a_i))+\sum(1+size(p_i))+\size(n)+\size(k)+1$. Note that
not all such codes represent instances of $(P'_k)$.
\end{example}
\noindent
The following decision procedure on an alternating Turing machine
generalises the $\mi{PSPACE}$ algorithms in~\cite{Vardi89}, given a
strictly one-step complete and contraction closed rule set~$\Rules$.
\begin{algorithm} \label{alg:sat}
(Decide satisfiability of $\phi\in \FLang(\ModSig)$)
\begin{enumerate}
\item \label{step:pseudoval}(Existential) Guess a propositionally
  consistent pseudovaluation~$H$ for $\phi$.
\item \label{step:clause} (Universal) Choose a contracted clause
  $\bot\neq\rho$ over $\ModAts(H)$ such that $H\PLentails\neg\rho$.
\item \label{step:matching} (Universal) Choose an $\RulesC$-matching
  $[R]$ of~$\rho$.
\item \label{step:premise} (Existential) Guess a clause~$\gamma$ from
  the CNF of the premise of~$R$.
\item \label{step:recursion} Recursively check that
  $\neg\gamma\sigma(R,\rho)$ is satisfiable.
\end{enumerate}
The algorithm succeeds if all possible choices at steps marked
\emph{universal} lead to successful termination, and for all steps
marked \emph{existential}, there exists a choice leading to successful
termination. Concerning Step~\ref{step:pseudoval}, note that the only
way for a pseudovaluation to be propositionally inconsistent is to
contain both $L\rho$ and $\neg L\rho$ for some modal atom $L\rho$.
\end{algorithm}
We emphasise that in Step~\ref{step:matching}, it suffices to guess one
code for each matching.
\begin{proposition}
Algorithm~\ref{alg:sat} succeeds iff the input formula~$\phi$ is
satisfiable.
\end{proposition}
\begin{proof}
Induction over the depth~$n$ of~$\phi$. If $n=0$, then the propositional
formula~$\phi$ will evaluate to either $\top$ or $\bot$, as it does not
contain any propositional variables; moreover, the only candidate for a
pseudovaluation for $\phi$ is the empty conjunctive clause $\top$. Thus,
the algorithm terminates unsuccessfully in the existential
step~(\ref{step:pseudoval}) iff~$\phi$ evaluates to $\bot$, since $\top$
is a pseudovaluation for $\phi$ iff $\phi$ evaluates to $\top$.
Otherwise, the algorithm terminates successfully in the universal
step~(\ref{step:clause}), since the only clause $\rho$ over
$\ModAts(\top)=\emptyset$ such that $\top\PLentails\neg\rho$ is $\bot$.
For $n>0$, correctness of the algorithm follows from
Corollary~\ref{cor:tableau-contracted} and the inductive hypothesis: the
essential demands of~$\phi$ are the negated premises
$\neg\phi\sigma(\phi/\psi,\rho)$ for $\RulesC$-matchings $[\phi/\psi]$
of contracted clauses~$\rho$ as in the algorithm, and such a demand is
satisfiable iff the negation of one of the clauses in the CNF of
$\phi\sigma(\phi/\psi,\rho)$ is satisfiable.
\end{proof}

\begin{remark}
In Step~\ref{step:pseudoval} of Algorithm~\ref{alg:sat}, it suffices to
consider the conjunctive clauses in some DNF of $\phi$ rather than all
pseudovaluations. A canonical, if not necessarily the most effective
choice for such a DNF is to take all pseudovaluations $H$ for $\phi$
such that $\ModAts(H)=\ModAts(\phi)$ (rather than only
$\ModAts(H)\subseteq\ModAts(\phi)$); in a concrete implementation, a
heuristic procedure for determining some DNF effectively may be
preferable.
\end{remark}

Note that due to the non-deterministic nature of the algorithm, the
above proposition does \emph{not} imply decidability of
$\FLang(\ModSig)$. This follows only if the algorithm respects suitable
resource bounds. We are interested in cases where the algorithm runs in
polynomial time. The crucial requirement for this is that
Steps~\ref{step:matching} and~\ref{step:premise} can be performed in
polynomial time, i.e.\ by suitable nondeterministic polynomial-time
multivalued functions (NPMV)~\cite{book-long-selman:npmv}. We recall
that a function $f: \Sigma^* \to \Pow{\Delta^*}$, where~$\Sigma$ and
$\Delta$ are alphabets, is NPMV iff
\begin{enumerate}
\item[(NPMV1)] there exists a polynomial~$p$ such that $|y| \leq p(|x|)$ for
all $y \in f(x)$, where~$|\cdot|$ denotes size, and
\item[(NPMV2)] the graph $\lbrace (x, y) \mid y \in f(x) \rbrace$ of~$f$ is in
  $\NP$.
\end{enumerate}
This motivates the following conditions:
\begin{definition}\label{def:tractable}
A set~$\Rules$ of rules is called \emph{$\PSPACE$-tractable} if
there exists a polynomial~$p$ such that all $\Rules$-matchings of a
contracted clause~$\rho$ over $\FLang(\ModSig)$ have some code of size at
most $p(|\rho|)$ (recall that matchings are equivalence classes of rules
and thus may have several codes), and it can be decided in $\NP$
\begin{enumerate}
\item\label{item:rule-np} whether a given code is the code of some rule in
  $\Rules$;
\item\label{item:match-np} whether a rule matches a given contracted clause; and 
\item\label{item:cnf-np} whether a clause belongs to the CNF of the
  premise of a given rule.
\end{enumerate}
\end{definition}
\begin{theorem}[Space Complexity] \label{thm:space}
Let~$\Rules$ be strictly one-step complete, closed under contraction,
and $\PSPACE$-tractable. Then the satisfiability problem for
$\FLang(\ModSig)$ is in $\PSPACE$.
\end{theorem}
\begin{proof}
Since~$\Rules$ is $\PSPACE$-tractable, so is~$\RulesC$, assuming
reasonable codes for the congruence rules (e.g.\ consisting of the
representation of the relevant modal operator; cf.\
Assumption~\ref{ass:formula-rep}). Thus, the functions mapping a clause
$\rho$ to the set of its $\RulesC$-matchings and a rule to the set of
clauses occurring in the CNF of its premise, respectively, are NPMV: in
the former case, the polynomial bound required by condition (NPMV1) is
ensured by the definition of $\PSPACE$-tractability, as we only need to
produce one code for each matching, and in the latter case, the
polynomial bound holds universally, as clauses are of polynomial
size. Condition (NPMV2) is ensured explicitly by
Definition~\ref{def:tractable} and Assumption~\ref{ass:formula-rep}
(which implies that the set of formulas is in $\NP$). Therefore,
Steps~\ref{step:matching} and~\ref{step:premise} in
Algorithm~\ref{alg:sat} can be performed in polynomial
time. Steps~\ref{step:pseudoval} and~\ref{step:clause} have polynomial
runtime without specific assumptions, as a pseudovaluation~$H$ for
$\phi$ is represented as a set of literals and must by definition
satisfy $\ModAts(H)\subseteq\ModAts(\phi)$, and the contracted clause
$\rho$ chosen in Step~\ref{step:clause} is constructed as a
non-repetitive list of literals whose negations belong to~$H$. Since the
depth of recursion is bounded by the depth of~$\phi$, it follows that
the algorithm runs in $\APTIME =
\PSPACE$~\cite{chandra-stockmeyer:aptime}.
\end{proof}

\begin{remark} A more careful
  analysis of Algorithm~\ref{alg:sat} reveals that it suffices for the
  decision problems in Definition~\ref{def:tractable} to be in
  $\mi{PH}$, the polynomial time hierarchy. In our examples, however,
  the complexity is in fact $P$ rather than $\NP$. We expect that
  this situation is typical, with the crucial condition for
  $\PSPACE$-tractability being the polynomial bound on
  $\Rules$-matchings. We are not aware of any natural examples of
  intractable rule sets (contrived examples are easy to construct, e.g.\
  by imposing computationally hard side conditions).
\end{remark}

\begin{remark} \label{rem:proof-search}
Algorithm \ref{alg:sat} can be dualised to yield a proof-search
procedure that determines whether $\phi \in \Lang(\Lambda)$ is
$\Rules$-derivable, thus implementing the shallow proof property
(Corollary~\ref{cor:shallow-proofs}/Theorem~\ref{thm:shallowproofs}). Note
that the dualisation entails that the roles of existential and universal
steps are interchanged.
\end{remark}

In the treatment of graded and propositional modal logic, the polynomial
bound on rule codes follows rather directly from size estimates in
integer linear programming, as follows. Following usual practice, we
take the \emph{size}~$|W|$ of a rational inequality $W\equiv(\sum_{i =
1}^n u_i x_i \mathrel\mathrm{op} u_0)$, $\mathrm{op} \in \lbrace <,
\leq, >, \geq \rbrace$ and $u_i \in \Rat$, to be $1+n+\sum_{i = 0}^n
\mi{size}(u_i)$.  We recall that for $n\in\Int$, $\sgn(n)=-1$ if $n<0$,
$\sgn(n)=1$ if $n>0$, and $\sgn(n)=0$ if $n=0$.

\begin{lemma}\label{lemma:poly}
For every rational linear inequality~$W$ and every solution $r_0, \dots,
r_n \in \Int$ of~$W$, there exists a solution $s_0, \dots,
s_n\in\Int$ of~$W$ such that $\sgn(s_i)=\sgn(r_i)$ for all~$i$,
the propositional formulas $\sum_{i=1}^n s_i a_i \geq s_0$ and
$\sum_{i=1}^n r_i a_i \geq r_0$ (cf.\ Section~\ref{sec:deduction}) are
equivalent, and $\size(s_i)\leq 18|W|^4$ for all~$i$.
\end{lemma}
\begin{proof}
Let $V = \lbrace a_1, \dots, a_n \rbrace$, and let $x_0,\dots,x_n$ be
the variables in~$W$. We note that a propositional formula $\sum_{i =
1}^n s_i a_i \geq s_0$ is equivalent to $\phi\equiv\sum_{i=1}^n r_i a_i
\geq a_0$ iff for all valuations $\nu: V \to \lbrace 0, 1 \rbrace$, one
has $\sum_{i=1}^n s_i \nu(a_i) \geq s_0$ if and only if $\sum_{i=1}^n
r_i \nu(a_i) \geq r_0$, read as integer linear inequalities. Thus, let
$I$ denote the system of inequalities consisting of~$W$ and additional
inqualities $F_i$ and~$E_\nu$, where $i=1,\dots,n$,~$\nu$ ranges over
valuations $V \to \lbrace 0, 1 \rbrace$,
\begin{equation*}
F_i=\begin{cases}x_i\ge 1 & \textrm{if $r_i\ge 1$}\\
x_i=0 & \textrm{if $r_i=0$}\\
               x_i\le -1 &\textrm{if $r_i\le-1$}\end{cases}
\end{equation*}
(where the middle case actually corresponds to two inequalities), and
\begin{equation*}
  E_\nu = \begin{cases}
  \sum_{i=1}^n x_i\nu(a_i) \geq x_0 & \mbox{if } \sum_{i=1}^n r_i
  \nu(a_i) \geq r_0 \\
  \sum_{i=1}^n x_i\nu(a_i) < x_0 & \mbox{if } \sum_{i=1}^n r_i
  \nu(a_i) < r_0.
  \end{cases}
\end{equation*}
Then the claim translates into the statement that~$I$ has a solution of
polynomially bounded size in~$|W|$.

It follows from \cite[Corollary 17.1b]{Schrijver86} that~$I$ has a
solution whose size is bounded by $6c (n+1)^3$, where~$c$ is the facet
complexity of the system, i.e.\ the size of the largest inequality in
$I$. As the cofficients of the inequalities~$E_\nu$ and~$F_i$ are of
size at most~$1$, we have $c\le|W|+2(n+1)$. Since moreover $|W|\ge n+1$,
$I$ thus has a solution of size at most $18|W|^4$.
\end{proof}
\noindent
We now illustrate how Theorem \ref{thm:space} allows us to establish
$\PSPACE$ bounds for many modal logics in a uniform way. 

\begin{example} \label{expl:main}
Conditions~(\ref{item:rule-np}) and~(\ref{item:match-np}) of
Definition~\ref{def:tractable} are immediate for all the rule sets of
Example~\ref{expl:resolution} --- the decision problems in question
involve no more than checking computationally harmless side conditions
in the case of Condition~(\ref{item:rule-np}) (disjointness and
containment of finite sets, linear inequalities), and comparing clauses
of polynomial (in fact, linear) size in the case of
Condition~(\ref{item:match-np}). Moreover, Condition~(\ref{item:cnf-np})
is immediate in those cases where the premises of rules are just single
clauses. This leaves only GML and PML; but the expansion of $\sum_{i\in
I}r_i a_i\ge k$ to a propositional formula is already in CNF, and checking whether a given clause
belongs to this CNF is clearly in~$P$.

It remains to establish the polynomial bound on the matchings. For GML
and PML, this is guaranteed precisely by Lemma~\ref{lemma:poly}. In all
other cases, every contracted clause~$\rho$ matches at most one rule,
whose code has size linear in the size of~$\rho$.

We thus have obtained $\PSPACE$-tractability and hence decidability in
$\PSPACE$ for all logics in Example~\ref{expl:resolution}. The logics
$E$ and~$M$ are of lesser interest here, being actually in
$\NP$~\cite{Vardi89}. We briefly comment on the algorithms and bounds
for the other cases.
\begin{longenum}
\item For the modal logics~$K$ and~$KD$
  (Examples~\ref{expl:resolution}.\ref{expl:k-osc} and~\ref{expl:kd-osc}),
  Algorithm~\ref{alg:sat} is essentially the witness
  algorithm~\cite{Ladner77,Vardi89,BlackburnEA01}. Both logics are
  $\PSPACE$-hard~\cite{Ladner77}.
\item For coalition logic
  (Example~\ref{expl:resolution}.\ref{expl:cml-osc}), we arrive, due to
  minor differences of the rule sets, at a slight variant of Pauly's
  $\PSPACE$-algorithm~\cite{Pauly02}.
\item For graded modal logic, we obtain a new algorithm which confirms
  the known $\PSPACE$ upper bound~\cite{Tobies01}. One might claim that
  the new algorithm is not only nicely embedded into a unified
  framework, but also conceptually simpler than the constraint-based
  algorithm of~\cite{Tobies01} (which corrects a similar but incorrect
  algorithm previously given elsewhere, and refutes a previous
  $\mi{EXPTIME}$ hardness conjecture). Graded modal logic is
  $\PSPACE$-hard, as it extends~$K$.
\item For probabilistic modal logic, we obtain a new algorithm which
  confirms the $\PSPACE$ upper bound that follows from the corresponding
  bound for the more expressive (modal) logic of probability, a proof of
  which is sketched in~\cite{FaginHalpern94}. The bound is tight, as PML
  contains the $\PSPACE$-complete logic~$KD$ as a fragment (embedded by
  mapping $\Box$ to~$L_1$). In comparison to the algorithm in
  \emph{loc.\ cit.}, our algorithm has additional proof theoretic
  content as discussed in Section~\ref{sec:shallowproofs}. Under the
  correspondence outlined in Remark~\ref{rem:proof-search}, it
  finds proofs which remain within PML rather than possibly diverting via
  a more expressive logic.
\item Our $\PSPACE$ upper bound for majority logic, which appeared for
  the first time in the conference presentation
  of~\cite{SchroderPattinson06}, tied in a priority race
  with~\cite{DemriLugiez06}, where a $\PSPACE$ upper bound was proved
  for the more expressive Presburger modal logic using a different type
  of algorithm. The same remarks concerning proof-theoretic content
  apply as for probabilistic modal logic.
\end{longenum}
\end{example} 

\section{Conclusion}

\noindent
Generalising results by~\citeN{Vardi89}, we have shown that coalgebraic
modal logic has the shallow model property, and we have presented a
generic $\PSPACE$ algorithm for satisfiability based on depth-first
exploration of shallow models. We have thus 
\begin{itemize}
\item reproduced the \emph{witness algorithm} for
 $K$ and~$KD$~\cite{BlackburnEA01}
\item obtained a slight variant of the known $\PSPACE$ algorithm for
  coalition logic~\cite{Pauly02}
\item obtained a new $\PSPACE$ algorithm for graded modal logic,
  recovering the known $\PSPACE$ bound~\cite{Tobies01}
\item obtained a new $\PSPACE$ algorithm for probabilistic modal
  logic~\cite{LarsenSkou91,HeifetzMongin01}, recovering a $\PSPACE$
  upper bound which follows from results sketched
  in~\cite{FaginHalpern94}.
\item obtained, simultaneously with~\cite{DemriLugiez06}, a new
  $\PSPACE$ upper bound for majority logic~\cite{PacuitSalame04}. 
\end{itemize}
In all these cases, the $\PSPACE$ upper bound is tight. Our algorithm
may alternatively be viewed as traversing a shallow proof that witnesses
a weak subformula property.

The crucial prerequisite for the generic algorithm is an axiomatisation
by so-called one-step rules (going from rank~$0$ to rank~$1$) obeying
two closedness conditions: closedness under resolution and under
contraction, i.e.\ removal of duplicate literals. In the examples, it
has not only turned out that it is feasible to keep this closure process
under control, but also that the axiomatisations obtained have
pleasingly compact presentations --- typically, one ends up with a single
rule schema. 

It has been shown that every modal logic can be equipped with a
canonical coalgebraic semantics, provided it is axiomatisable in
rank~$1$ and satisfies the congruence
rule~\cite{SchroderPattinson07mcs}. This means in particular that our
shallow model construction applies to every such modal logic when
equipped with the canonical semantics. Moreover, the $\PSPACE$-algorithm
presented here can be made modular w.r.t.\ heterogeneous combination of
systems and modal logics using multi-sorted
coalgebra~\cite{SchroderPattinson07}. The extension of the theory beyond
rank~$1$ is the subject of future research, as is the treatment of
simple fixed point operators, possibly using automata theoretic
methods~\cite{Vardi96,Venema06} or pseudomodels~\cite{EmersonHalpern85}.
A further point of interest is to investigate the connection between our
notion of resolution closure and classical proof-theoretic issues such
as cut elimination and interpolation.

\begin{acks}  The authors wish to thank Alexander Kurz for
useful discussions
and the Department of Computer Science at the University of
Bremen for funding a visit of the second author.
\end{acks}

\bibliographystyle{acmtrans}
\bibliography{coalgml}

\begin{thebibliography}{}

\bibitem[\protect\citeauthoryear{Barr}{Barr}{1993}]{Barr93}
{\sc Barr, M.} 1993.
\newblock Terminal coalgebras in well-founded set theory.
\newblock {\em Theor. Comput. Sci.\/}~{\em 114}, 299--315.

\bibitem[\protect\citeauthoryear{Bartels}{Bartels}{2003}]{Bartels03}
{\sc Bartels, F.} 2003.
\newblock Generalised coinduction.
\newblock {\em Math. Struct. Comput. Sci.\/}~{\em 13}, 321--348.

\bibitem[\protect\citeauthoryear{Bartels, Sokolova, and de~Vink}{Bartels
  et~al\mbox{.}}{2004}]{BartelsEA04}
{\sc Bartels, F.}, {\sc Sokolova, A.}, {\sc and} {\sc de~Vink, E.~P.} 2004.
\newblock A hierarchy of probabilistic system types.
\newblock {\em Theor. Comput. Sci.\/}~{\em 327}, 3--22.

\bibitem[\protect\citeauthoryear{Blackburn, de~Rijke, and Venema}{Blackburn
  et~al\mbox{.}}{2001}]{BlackburnEA01}
{\sc Blackburn, P.}, {\sc de~Rijke, M.}, {\sc and} {\sc Venema, Y.} 2001.
\newblock {\em Modal Logic}.
\newblock Cambridge University Press, Cambridge.

\bibitem[\protect\citeauthoryear{Book, Long, and Selman}{Book
  et~al\mbox{.}}{1984}]{book-long-selman:npmv}
{\sc Book, R.}, {\sc Long, T.}, {\sc and} {\sc Selman, A.} 1984.
\newblock Quantitative relativizations of complexity classes.
\newblock {\em SIAM J. Comput.\/}~{\em 13}, 461--487.

\bibitem[\protect\citeauthoryear{Carlyle and Paz}{Carlyle and
  Paz}{1971}]{CarlylePaz71}
{\sc Carlyle, J.~W.} {\sc and} {\sc Paz, A.} 1971.
\newblock Realizations by stochastic finite automata.
\newblock {\em J. Comput. Syst. Sci.\/}~{\em 5}, 26--40.

\bibitem[\protect\citeauthoryear{Caro}{Caro}{1988}]{Caro88}
{\sc Caro, F.~D.} 1988.
\newblock Graded modalities {II} (canonical models).
\newblock {\em Studia logica\/}~{\em 47}, 1--10.

\bibitem[\protect\citeauthoryear{Chandra, Kozen, and Stockmeyer}{Chandra
  et~al\mbox{.}}{1981}]{chandra-stockmeyer:aptime}
{\sc Chandra, A.}, {\sc Kozen, D.}, {\sc and} {\sc Stockmeyer, L.} 1981.
\newblock Alternation.
\newblock {\em J.\ ACM\/}~{\em 28}, 114--133.

\bibitem[\protect\citeauthoryear{Chellas}{Chellas}{1980}]{Chellas80}
{\sc Chellas, B.} 1980.
\newblock {\em Modal Logic}.
\newblock Cambridge University Press, Cambridge.

\bibitem[\protect\citeauthoryear{C{\^i}rstea and Pattinson}{C{\^i}rstea and
  Pattinson}{2007}]{CirsteaPattinson07}
{\sc C{\^i}rstea, C.} {\sc and} {\sc Pattinson, D.} 2007.
\newblock Modular construction of complete coalgebraic logics.
\newblock {\em Theor. Comput. Sci.\/}.
\newblock In press.

\bibitem[\protect\citeauthoryear{D'Agostino and Visser}{D'Agostino and
  Visser}{2002}]{DAgostinoVisser02}
{\sc D'Agostino, G.} {\sc and} {\sc Visser, A.} 2002.
\newblock Finality regained: A coalgebraic study of {Scott}-sets and multisets.
\newblock {\em Arch.\ Math.\ Logic\/}~{\em 41}, 267--298.

\bibitem[\protect\citeauthoryear{De~Nivelle, Schmidt, and Hustadt}{De~Nivelle
  et~al\mbox{.}}{2000}]{DeNivelleEA00}
{\sc De~Nivelle, H.}, {\sc Schmidt, R.~A.}, {\sc and} {\sc Hustadt, U.} 2000.
\newblock Resolution-based methods for modal logics.
\newblock {\em Logic J.\ IGPL\/}~{\em 8}, 265--292.

\bibitem[\protect\citeauthoryear{Demri and Lugiez}{Demri and
  Lugiez}{2006}]{DemriLugiez06}
{\sc Demri, S.} {\sc and} {\sc Lugiez, D.} 2006.
\newblock {P}resburger modal logic is only {PSPACE}-complete.
\newblock In {\em {IJCAR} 2006, {P}roceedings of the Third {I}nternational
  {J}oint {C}onference on {A}utomated {R}easoning}, {U.~Furbach} {and}
  {N.~Shankar}, Eds. Lect.\ Notes Artificial Intell., vol. 4130. Springer,
  Berlin, 541--556.
\newblock Full version available as Research Report LSV-06-15, Laboratoire
  Sp{\'e}cification et V{\'e}rification, Ecole Normale Sup{\'e}rieure de
  Cachan, 2006.

\bibitem[\protect\citeauthoryear{Emerson and Halpern}{Emerson and
  Halpern}{1985}]{EmersonHalpern85}
{\sc Emerson, E.~A.} {\sc and} {\sc Halpern, J.~Y.} 1985.
\newblock Decision procedures and expressiveness in the temporal logic of
  branching time.
\newblock {\em J. Comput. Syst. Sci.\/}~{\em 30}, 1--24.

\bibitem[\protect\citeauthoryear{Fagin and Halpern}{Fagin and
  Halpern}{1994}]{FaginHalpern94}
{\sc Fagin, R.} {\sc and} {\sc Halpern, J.~Y.} 1994.
\newblock Reasoning about knowledge and probability.
\newblock {\em J.\ ACM\/}~{\em 41}, 340--367.

\bibitem[\protect\citeauthoryear{Fine}{Fine}{1972}]{Fine72}
{\sc Fine, K.} 1972.
\newblock In so many possible worlds.
\newblock {\em Notre Dame J.\ Formal Logic\/}~{\em 13}, 516--520.

\bibitem[\protect\citeauthoryear{Halpern and R{\^e}go}{Halpern and
  R{\^e}go}{2007}]{HalpernRego06}
{\sc Halpern, J.} {\sc and} {\sc R{\^e}go, L.~C.} 2007.
\newblock Characterizing the {NP-PSPACE} gap in the satisfiability problem for
  modal logic.
\newblock In {\em {IJCAI} 2007, Proceedings of the 20th International Joint
  Conference on Artificial Intelligence}, {M.~M. Veloso}, Ed. 2306--2311.

\bibitem[\protect\citeauthoryear{Hansen and Kupke}{Hansen and
  Kupke}{2004}]{HansenKupke04}
{\sc Hansen, H.~H.} {\sc and} {\sc Kupke, C.} 2004.
\newblock A coalgebraic perspective on monotone modal logic.
\newblock In {\em Coalgebraic Methods in Computer Science}, {J.~Ad{\'a}mek}
  {and} {S.~Milius}, Eds. Electron.\ Notes Theor.\ Comput.\ Sci., vol. 106.
  Elsevier, Amsterdam, 121--143.

\bibitem[\protect\citeauthoryear{Heifetz and Mongin}{Heifetz and
  Mongin}{2001}]{HeifetzMongin01}
{\sc Heifetz, A.} {\sc and} {\sc Mongin, P.} 2001.
\newblock Probabilistic logic for type spaces.
\newblock {\em Games and Economic Behavior\/}~{\em 35}, 31--53.

\bibitem[\protect\citeauthoryear{Jacobs}{Jacobs}{2000}]{Jacobs00}
{\sc Jacobs, B.} 2000.
\newblock Towards a duality result in coalgebraic modal logic.
\newblock In {\em CMCS 2000, Coalgebraic Methods in Computer Science},
  {H.~Reichel}, Ed. Electron.\ Notes Theor.\ Comput.\ Sci., vol.~33. Elsevier,
  Amsterdam.

\bibitem[\protect\citeauthoryear{Kupke, Kurz, and Pattinson}{Kupke
  et~al\mbox{.}}{2005}]{KupkeEA05}
{\sc Kupke, C.}, {\sc Kurz, A.}, {\sc and} {\sc Pattinson, D.} 2005.
\newblock Ultrafilter extensions for coalgebras.
\newblock In {\em {CALCO} 2005, Algebra and Coalgebra in Computer Science:
  First International Conference, Proceedings}, {J.~L. Fiadeiro}, {N.~Harman},
  {M.~Roggenbach}, {and} {J.~Rutten}, Eds. Lect.\ Notes Comput.\ Sci., vol.
  3629. Springer, Berlin, 263--277.

\bibitem[\protect\citeauthoryear{Kurz}{Kurz}{2001}]{Kurz01}
{\sc Kurz, A.} 2001.
\newblock Specifying coalgebras with modal logic.
\newblock {\em Theor. Comput. Sci.\/}~{\em 260}, 119--138.

\bibitem[\protect\citeauthoryear{Ladner}{Ladner}{1977}]{Ladner77}
{\sc Ladner, R.} 1977.
\newblock The computational complexity of provability in systems of modal
  propositional logic.
\newblock {\em SIAM J.\ Comput.\/}~{\em 6}, 467--480.

\bibitem[\protect\citeauthoryear{Larsen and Skou}{Larsen and
  Skou}{1991}]{LarsenSkou91}
{\sc Larsen, K.} {\sc and} {\sc Skou, A.} 1991.
\newblock Bisimulation through probabilistic testing.
\newblock {\em Inf. Comput.\/}~{\em 94}, 1--28.

\bibitem[\protect\citeauthoryear{Mossakowski, Schr{\"o}der, Roggenbach, and
  Reichel}{Mossakowski et~al\mbox{.}}{2006}]{MossakowskiEA04}
{\sc Mossakowski, T.}, {\sc Schr{\"o}der, L.}, {\sc Roggenbach, M.}, {\sc and}
  {\sc Reichel, H.} 2006.
\newblock Algebraic-coalgebraic specification in {\CoCASL}.
\newblock {\em J. Logic Algebraic Programming\/}~{\em 67}, 146--197.

\bibitem[\protect\citeauthoryear{Ohlbach and Koehler}{Ohlbach and
  Koehler}{1999}]{OhlbachKoehler99}
{\sc Ohlbach, H.~J.} {\sc and} {\sc Koehler, J.} 1999.
\newblock Modal logics, description logics and arithmetic reasoning.
\newblock {\em Artificial Intelligence\/}~{\em 109}, 1--31.

\bibitem[\protect\citeauthoryear{Pacuit and Salame}{Pacuit and
  Salame}{2004}]{PacuitSalame04}
{\sc Pacuit, E.} {\sc and} {\sc Salame, S.} 2004.
\newblock Majority logic.
\newblock In {\em KR 2004, Principles of Knowledge Representation and
  Reasoning: Proceedings of the Ninth International Conference}, {D.~Dubois},
  {C.~A. Welty}, {and} {M.-A. Williams}, Eds. AAAI Press, 598--605.

\bibitem[\protect\citeauthoryear{Pattinson}{Pattinson}{2001}]{Pattinson01}
{\sc Pattinson, D.} 2001.
\newblock Semantical principles in the modal logic of coalgebras.
\newblock In {\em {STACS} 2001, 18th Annual Symposium on Theoretical Aspects of
  Computer Science, Proceedings}, {A.~Ferreira} {and} {H.~Reichel}, Eds. Lect.\
  Notes Comput.\ Sci., vol. 2010. Springer, Berlin, 514--526.

\bibitem[\protect\citeauthoryear{Pattinson}{Pattinson}{2003}]{Pattinson03}
{\sc Pattinson, D.} 2003.
\newblock Coalgebraic modal logic: Soundness, completeness and decidability of
  local consequence.
\newblock {\em Theor. Comput. Sci.\/}~{\em 309}, 177--193.

\bibitem[\protect\citeauthoryear{Pattinson}{Pattinson}{2004}]{Pattinson04}
{\sc Pattinson, D.} 2004.
\newblock Expressive logics for coalgebras via terminal sequence induction.
\newblock {\em Notre Dame J.\ Formal Logic\/}~{\em 45}, 19--33.

\bibitem[\protect\citeauthoryear{Pauly}{Pauly}{2002}]{Pauly02}
{\sc Pauly, M.} 2002.
\newblock A modal logic for coalitional power in games.
\newblock {\em J. Logic and Comput.\/}~{\em 12}, 149--166.

\bibitem[\protect\citeauthoryear{Pauly}{Pauly}{2005}]{Pauly05}
{\sc Pauly, M.} 2005.
\newblock On the role of language in social choice theory.
\newblock Unpublished manuscript.

\bibitem[\protect\citeauthoryear{Rabin}{Rabin}{1963}]{Rabin63}
{\sc Rabin, M.} 1963.
\newblock Probabilistic automata.
\newblock {\em Inform.\ Control\/}~{\em 6}, 230--245.

\bibitem[\protect\citeauthoryear{R{\"o\ss}iger}{R{\"o\ss}iger}{2000}]{Roessige%
r00}
{\sc R{\"o\ss}iger, M.} 2000.
\newblock Coalgebras and modal logic.
\newblock In {\em CMCS 2000, Coalgebraic Methods in Computer Science},
  {H.~Reichel}, Ed. Electron.\ Notes Theor.\ Comput.\ Sci., vol.~33. Elsevier,
  Amsterdam.

\bibitem[\protect\citeauthoryear{Rothe, Tews, and Jacobs}{Rothe
  et~al\mbox{.}}{2001}]{RotheTewsJacobs01}
{\sc Rothe, J.}, {\sc Tews, H.}, {\sc and} {\sc Jacobs, B.} 2001.
\newblock The {C}oalgebraic {C}lass {S}pecification {L}anguage {CCSL}.
\newblock {\em J. Universal Comput. Sci.\/}~{\em 7}, 175--193.

\bibitem[\protect\citeauthoryear{Rutten}{Rutten}{2000}]{Rutten00}
{\sc Rutten, J.} 2000.
\newblock Universal coalgebra: A theory of systems.
\newblock {\em Theor. Comput. Sci.\/}~{\em 249}, 3--80.

\bibitem[\protect\citeauthoryear{Schrijver}{Schrijver}{1986}]{Schrijver86}
{\sc Schrijver, A.} 1986.
\newblock {\em Theory of linear and integer programming}.
\newblock John Wiley \& Sons, Chichester.

\bibitem[\protect\citeauthoryear{Schr{\"o}der}{Schr{\"o}der}{2005}]{Schroder05}
{\sc Schr{\"o}der, L.} 2005.
\newblock Expressivity of coalgebraic modal logic: the limits and beyond.
\newblock In {\em {FOSSACS} 2005, Foundations of Software Science and
  Computation Structures, 8th International Conference, Proceedings},
  {V.~Sassone}, Ed. Lect.\ Notes Comput.\ Sci., vol. 3441. Springer, Berlin,
  440--454.
\newblock Extended version to appear in {\emph{Theor.\ Comput.\ Sci.}}

\bibitem[\protect\citeauthoryear{Schr{\"o}der}{Schr{\"o}der}{2007}]{Schroder06}
{\sc Schr{\"o}der, L.} 2007.
\newblock A finite model construction for coalgebraic modal logic.
\newblock {\em J.\ Logic Algebraic Programming\/}.
\newblock In press. Earlier version in {\emph{Foundations of Software Science
  And Computation Structures}}, vol.\ 3921 of Lect.\ Notes Comput.\ Sci., pp.\
  157--171, Springer, Berlin, 2006.

\bibitem[\protect\citeauthoryear{Schr{\"{o}}der and Pattinson}{Schr{\"{o}}der
  and Pattinson}{2006}]{SchroderPattinson06}
{\sc Schr{\"{o}}der, L.} {\sc and} {\sc Pattinson, D.} 2006.
\newblock {PSPACE} reasoning for rank-1 modal logics.
\newblock In {\em LICS 2006, Proceedings of the 21st Annual IEEE Symposium on
  Logic in Computer Science}, {R.~Alur}, Ed. IEEE Computer Society Press,
  231--240.
\newblock Presentation slides available under
  www.informatik.uni-bremen.de/$\sim$lschrode/slides/rank1pspace.pdf.

\bibitem[\protect\citeauthoryear{Schr{\"o}der and Pattinson}{Schr{\"o}der and
  Pattinson}{2007a}]{SchroderPattinson07}
{\sc Schr{\"o}der, L.} {\sc and} {\sc Pattinson, D.} 2007a.
\newblock Modular algorithms for heterogeneous modal logics.
\newblock In {\em ICALP 2007, Automata, Languages and Programming, 34th
  International Colloquium, Proceedings}, {L.~Age}, {A.~Tarlecki}, {and}
  {C.~Cachin}, Eds. Lect.\ Notes Comput.\ Sci. Springer, Berlin.
\newblock To appear.

\bibitem[\protect\citeauthoryear{Schr{\"o}der and Pattinson}{Schr{\"o}der and
  Pattinson}{2007b}]{SchroderPattinson07mcs}
{\sc Schr{\"o}der, L.} {\sc and} {\sc Pattinson, D.} 2007b.
\newblock Rank-1 modal logics are coalgebraic.
\newblock In {\em {STACS} 2007, 24th Annual Symposium on Theoretical Aspects of
  Computer Science, Proceedings}, {W.~Thomas} {and} {P.~Weil}, Eds. Lect.\
  Notes Comput.\ Sci., vol. 4393. Springer, Berlin, 574--585.

\bibitem[\protect\citeauthoryear{Tobies}{Tobies}{2001}]{Tobies01}
{\sc Tobies, S.} 2001.
\newblock {$\mi{PSPACE}$} reasoning for graded modal logics.
\newblock {\em J. Logic and Comput.\/}~{\em 11}, 85--106.

\bibitem[\protect\citeauthoryear{Troelstra and Schwichtenberg}{Troelstra and
  Schwichtenberg}{1996}]{TroelstraSchwichtenberg96}
{\sc Troelstra, A.~S.} {\sc and} {\sc Schwichtenberg, H.} 1996.
\newblock {\em Basic Proof Theory}. Cambridge Tracts in Theoretical Computer
  Science, vol.~43.
\newblock Cambridge University Press, Cambridge.

\bibitem[\protect\citeauthoryear{Turi and Plotkin}{Turi and
  Plotkin}{1997}]{TuriPlotkin97}
{\sc Turi, D.} {\sc and} {\sc Plotkin, G.} 1997.
\newblock Towards a mathematical operational semantics.
\newblock In {\em LICS 1997, Proceedings of the 12th Annual IEEE Symposium on
  Logic in Computer Science}. IEEE Computer Society Press, 280--291.

\bibitem[\protect\citeauthoryear{van~der Hoek and Meyer}{van~der Hoek and
  Meyer}{1992}]{vdHoekMeyer92}
{\sc van~der Hoek, W.} {\sc and} {\sc Meyer, J.-J.} 1992.
\newblock Graded modalities in epistemic logic.
\newblock In {\em LFCS 1992, Logical Foundations of Computer Science, Second
  International Symposium, Proceedings}, {A.~Nerode} {and} {M.~A. Taitslin},
  Eds. Lect.\ Notes Comput.\ Sci., vol. 620. Springer, Berlin, 503--514.

\bibitem[\protect\citeauthoryear{Vardi}{Vardi}{1989}]{Vardi89}
{\sc Vardi, M.} 1989.
\newblock On the complexity of epistemic reasoning.
\newblock In {\em LICS 1989, Proceedings of the Fourth Annual IEEE Symposium on
  Logic in Computer Science}. IEEE Computer Society Press, 243--251.

\bibitem[\protect\citeauthoryear{Vardi}{Vardi}{1996}]{Vardi96}
{\sc Vardi, M.~Y.} 1996.
\newblock Why is modal logic so robustly decidable?
\newblock In {\em Descriptive Complexity and Finite Models, Proceedings of a
  {DIMACS} Workshop}, {N.~Immerman} {and} {P.~G. Kolaitis}, Eds. DIMACS Ser.\
  in Discrete Math.\ and Theor.\ Comput.\ Sci., vol.~31. American Mathematical
  Society, 149--184.

\bibitem[\protect\citeauthoryear{Venema}{Venema}{2006}]{Venema06}
{\sc Venema, Y.} 2006.
\newblock Automata and fixed point logics: a coalgebraic perspective.
\newblock {\em Inf. Comput.\/}~{\em 204}, 637--678.

\end{thebibliography}

\end{document}